\documentclass[twocolumn]{aastex62}
\usepackage{float}
\usepackage{textcomp}
\usepackage{graphicx}
\usepackage[caption=false]{subfig}
\usepackage{enumitem}
\usepackage{comment}
\usepackage{makecell}
\usepackage{environ}
\usepackage{multirow}
\usepackage{xurl}
\usepackage{amsmath}
\usepackage{orcidlink}

\usepackage{lineno}

\received{X Y, ZZZZ}
\revised{X Y, ZZZZ}
\accepted{\today}
\submitjournal{ApJ}

\begin{document}
\title{Search for Kink Events in Variable \textit{Fermi}-LAT blazars}

\correspondingauthor{P. Peñil, H. Zhang, J. Otero-Santos}
\email{E-mails: ppenil@clemson.edu, haocheng.zhang@nasa.gov}
\email{jorge.otero@pd.infn.it}

\author{P. Pe\~nil\orcidlink{0000-0003-3741-9764}}
\affil{Department of Physics and Astronomy, Clemson University,	Kinard Lab of Physics, Clemson, SC 29634-0978, USA}

\author{H. Zhang\orcidlink{0000-0001-9826-1759}}
\affil{University of Maryland Baltimore County, Baltimore, MD 21250, USA}
\affil{NASA Goddard Space Flight Center, Greenbelt, MD 20771, USA}

\author{J. Otero-Santos\orcidlink{0000-0002-4241-5875}}
\affil{Istituto Nazionale di Fisica Nucleare, Sezione di Padova, 35131 Padova, Italy}

\author{M. Ajello\orcidlink{0000-0002-6584-1703}}
\affil{Department of Physics and Astronomy, Clemson University,	Kinard Lab of Physics, Clemson, SC 29634-0978, USA}

\author{S. Buson\orcidlink{0000-0002-3308-324X}}
\affil{Julius-Maximilians-Universit\"at W\"urzburg, Fakultät f\"ur Physik und Astronomie, Emil-Fischer-Str. 31, D-97074 W\"urzburg, Germany}
\affil{Deutsches Elektronen-Synchrotron DESY, Platanenallee 6, 15738 Zeuthen, Germany}

\author{S. Adhikari\orcidlink{0009-0006-1029-1026}}
\affil{Department of Physics and Astronomy, Clemson University,	Kinard Lab of Physics, Clemson, SC 29634-0978, USA}

\author{A. Rico\orcidlink{0000-0001-5233-7180}}
\affil{Department of Physics and Astronomy, Clemson University,	Kinard Lab of Physics, Clemson, SC 29634-0978, USA}

\author{J. Escudero Pedrosa\orcidlink{0000-0002-4131-655X}}
\affil{Center for Astrophysics | Harvard \& Smithsonian, Cambridge, MA 02138, USA}

\author{I. Agudo\orcidlink{0000-0002-3777-6182}}
\affil{Instituto de Astrof\'isica de Andalucía (CSIC), Glorieta de la Astronomía s/n, 18008 Granada, Spain}

\author{D. Morcuende\orcidlink{0000-0001-9400-0922}}
\affil{Instituto de Astrof\'isica de Andalucía (CSIC), Glorieta de la Astronomía s/n, 18008 Granada, Spain}

\author{A. Sota\orcidlink{0000-0002-9404-6952}}
\affil{Instituto de Astrof\'isica de Andalucía (CSIC), Glorieta de la Astronomía s/n, 18008 Granada, Spain}

\author{V. Casanova\orcidlink{0000-0003-2036-8999}}
\affil{Instituto de Astrof\'isica de Andalucía (CSIC), Glorieta de la Astronomía s/n, 18008 Granada, Spain}

\author{F. J. Aceituno\orcidlink{0000-0001-8074-4760}}
\affil{Instituto de Astrof\'isica de Andalucía (CSIC), Glorieta de la Astronomía s/n, 18008 Granada, Spain}

\begin{abstract}
This study explores the detection of Quasi-Periodic Oscillations (QPOs) in blazars as a method to identify kink events within their jets, utilizing both $\gamma$-ray and polarized light observations. Focusing on a sample of 9 blazars, we analyze $\gamma$-ray light curves to identify significant QPOs. In addition to $\gamma$-ray data, we incorporated polarized light data corresponding to the same temporal segments to cross-validate the presence of QPOs. However, the limited availability of comprehensive polarized data restricted our ability to perform a thorough analysis across all datasets. Despite these limitations, our analysis reveals a segment where QPOs in polarized light coincided with those observed in $\gamma$-rays, providing preliminary evidence supporting the kink origin of these oscillations.

\end{abstract}

\keywords{BL Lacertae objects, methods: data analysis, telescopes: Fermi-LAT}

\section{Introduction}\label{sec:intro}

Blazars are active galaxies hosting relativistic plasma jets arising from supermassive black holes that point very close to our line of sight. Due to the relativistic beaming effects, they are the most numerous extragalactic $\gamma$-ray sources detected by \textit{Fermi} Large Area Telescope \citep[\textit{Fermi}-LAT,][]{fermi_lat}. They exhibit highly variable nonthermal-dominated emission across the entire electromagnetic spectrum. The variable multi-wavelength emission comes from an unresolved region, often referred to as the blazar zone, that locates somewhere between sub-parsec to a few parsecs from the central engine. The GeV to TeV $\gamma$-rays can flare in as short as a few minutes as observed by \textit{Fermi}-LAT and ground-based TeV air Cherenkov telescopes \citep{Ackermann2016,Aharonian2007,Albert2007}, indicating extremely efficient particle acceleration within the blazar zone. The low-energy spectral component, ranging from radio to optical-UV, in some blazars up to X-rays, originates from synchrotron emission by nonthermal electrons in a partially ordered magnetic field. This is evident by the high polarization observed in radio to optical, and recently in X-rays for some blazars \citep{Scarpa1997,Lister2018,Liodakis2022}. The high-energy emission may have a hadronic contribution, as suggested by the recent detection of the contemporaneous IceCube neutrino and \textit{Fermi}-LAT flare event in the blazar TXS~0506+056 \citep{IceCube2018}. However, in most cases the X-ray to $\gamma$-ray emission is well modeled by inverse Compton scattering by the same electrons that synchrotron emit the radio to optical spectral component. Under this scenario, the $\gtrsim$GeV $\gamma$-rays observed by \textit{Fermi}-LAT are coming from the roughly same electrons that produce the optical emission. It is generally believed that the blazar jet is highly magnetized at the jet launching site \citep{Blandford1977,Tchekhovskoy2010}. However, how the magnetic energy dissipates to drive the bulk relativistic motion of the jet and acceleration of nonthermal particles remains unclear. Given the strong particle acceleration in the blazar zone, it is often considered the main location where the bulk jet energy dissipates. But the energy partition at the blazar zone is also under debate. It is therefore important to understand how the magnetic field affects the jet propagation and interaction with the surrounding medium, which determine the locations of energy dissipation and the energy partition therein. In particular, we need to identify characteristic signatures that can reveal the jet dynamics and energy partition.

Kink instabilities are a type of current-driven magnetohydrodynamic (MHD) fluid instabilities. They can happen in a blazar jet where the magnetic energy takes a significant portion of the total jet energy and the toroidal magnetic field component is considerable. Kink instabilities can cause transverse displacements of the jet plasma and twist the magnetic field into spiral-like structures. Earlier works show that in a relativistic jet pervaded by helical magnetic fields, kink instabilities can naturally develop and convert a significant jet magnetic energy into other forms, which can lead to the acceleration of nonthermal particles through magnetic reconnection and/or turbulence \citep{Alves2018,Davelaar2020,Begelman1998}. The spiral-like magnetic field and current density in the kinked region form a clear quasi-periodic structure. Each repeating quasi-periodic segment is often referred to as a kink node. Kink instabilities will eventually saturate and likely develop turbulence-like structures, making the quasi-periodic structure a transient physical phenomenon. In principle, kink instabilities can strongly twist the jet and eventually disrupt the jet propagation. But large-scale MHD simulations of magnetic-driven jets have often shown that, very interestingly, kink instabilities mostly affect the central spine of the jet, but they do not disrupt the global jet structure and propagation \citep{Guan14,Barniol_Duran17}. This is very likely due to the returning poloidal magnetic field component that envelops the jet, which can stabilize the MHD instabilities towards the jet boundary. Strong kink instabilities in the central spine of the jet can be sites for strong magnetic energy dissipation and particle acceleration, a promising scenario for the blazar zone.

The characteristic magnetic field morphology in kink instabilities can leave unique observable signatures in polarization. \citet{haocheng_polarization_signatures} simulates local kink instabilities inside the blazar zone and finds that they can produce simultaneous variations in synchrotron flux and polarization. \citet{kink_qpos} performs the study in a global MHD jet simulation and, for the first time, shows that kink instabilities can drive simultaneous quasi-periodic oscillations (QPOs) in flux and polarization. Such QPOs originate from a part of or roughly one kink node that becomes very bright due to local physical conditions, such as shocks or changes in the density of ambient medium that enhance the local kink instabilities. Given that quasi-periodic structures generated by kink are transient phenomena, such simultaneous QPOs in flux and polarization are also transient and can disappear after a number of cycles. \citet{Acharya2021} confirms the QPOs in flux from one kink node based on a local kink instability simulation and shows that if multiple kink nodes are comparably bright, the QPOs from all kink nodes will be washed out, probably because the QPOs in each kink node are not synchronized in phase. This explains why \citet{haocheng_polarization_signatures} does not find QPOs since that work is focused on the radiation and polarization signatures from the entire segment of the kinked jet containing multiple kink nodes. These theoretical studies provide a novel method to identify kink instabilities in blazars via the detection of transient simultaneous QPOs in multi-wavelength flux and polarization signatures. If confirmed by observations, kink instabilities in the blazar zone unambiguously infer that a significant amount of the jet energy is still in the form of magnetic energy at the blazar zone, and magnetic reconnection and turbulence likely drive the particle acceleration there.

\citet{kink_bllacertae} reports the first and so far the only confirmed kink-driven transient QPOs in blazars. This paper presents the simultaneous long-term monitoring of optical flux and polarization of the blazar BL Lacertae, performed by the Whole Earth Blazar Telescope team. In August 2020 (starting from MJD 29076), during a flaring state, BL Lacertae made more than ten cycles of simultaneous QPOs in optical and $\gamma$-ray flux and optical polarization with a period of $\sim 0.55$ days, which can be well interpreted by the kink-driven transient QPOs. Interestingly, the paper also shows a curved radio jet image right after the QPO flaring event, further suggesting the presence of kink instabilities in the blazar. Transient QPOs in various observational bands have been reported in other blazars as well, with periods ranging from a few days to a few months \citep{Hayashida1998,Rani2009,Zhou2018}. Although some recent works suggest that they may be associated with the kink-driven QPOs \citep{Tripathi2024,Lu2024,Prince2023}, their results are unfortunately inconclusive mainly for three issues. Firstly, there is a limited amount of continuous optical polarization monitoring data. The highly variable nature of kink-driven QPOs in polarization demands high cadence and continuous optical polarization monitoring, which requires coordinated observational campaigns involving multiple optical polarimeters across the globe. Secondly, it is unclear how to trigger a dedicated multi-wavelength campaign for kink-driven QPOs. Since they are by nature transient, they rely on blazar monitoring telescopes such as \textit{Fermi}-LAT to trigger campaigns. However, it is not like a traditional trigger that is based on the flux level; it is a time-domain trigger that is based on a specific temporal pattern. Lastly, kink-driven QPOs lack standard analysis procedures. So far, we do not know how many cycles of transient QPOs are statistically significant to confirm kink instabilities and how gaps in multi-wavelength flux and polarization data and systematic uncertainties may affect the results. Additionally, accurately determining the presence of a QPO requires more than simply counting the number of observed cycles. A robust identification demands statistical evidence that effectively distinguishes the signal from a noise origin. It is crucial to account for the stochastic nature of potential QPOs, as random noise can produce spurious periodic patterns, potentially leading to false detection \citep[][]{vaughan_criticism}. These last issues are of particular importance as they also determine how much optical polarization data is needed.

This paper aims to establish a general analysis procedure for the kink-driven transient QPOs in blazars. We will show that the highly correlated flux and polarization QPOs are unlikely reproduced by red noises, but strongly favor a physical origin. Additionally, we find that only a few cycles are sufficient to confirm a kink-driven QPO event, ideal for dedicated multi-wavelength campaigns to probe such transient phenomena. The continuous $\gamma$-ray monitoring capability of \textit{Fermi}-LAT is crucial to trigger multi-wavelength observations with polarization to identify more kink-driven QPOs. The paper is organized as follows. $\S$\ref{sec:kink_model} describes a semi-analytical model for kink-driven QPOs, based on the MHD simulations and QPO models in \citet{kink_qpos}. Next, $\S$\ref{sec:test} describes various tests to enhance the reliability and robustness of our results. In $\S$\ref{sec:fermidata}, we present the data sample analyzed along with the multi-wavelength data collected. $\S$\ref{sec:methodology} outlines the periodicity analysis methodology. $\S$\ref{sec:results} provides the results of our study. Finally, we summarize our findings in $\S$\ref{sec:summary}.

\section{Kink-Driven QPO Model} \label{sec:kink_model}

Kink instabilities are complex plasma dynamical processes that are impossible to be fully described by semi-analytical formulae. Nonetheless, \citet{kink_qpos} and \citet{kink_bllacertae} have shown that the kink-driven QPOs can be approximated by a strictly periodic component plus a turbulent component. \citet{kink_qpos} have shown that if one or a part of a kink node becomes very bright due to interactions with surrounding medium, this kink node can quasi-periodically inject nonthermal electrons in its central spine, where the poloidal magnetic field dominates over the toroidal component. The resulting QPOs are not only in the total flux, but also have a preference in the polarization direction (electric vector polarization angle, EVPA), which is perpendicular to the jet direction. The period can be approximated by $T_1=\frac{R_{tr}}{v_{tr}\Gamma}$, where $R_{tr}$ is the cross section of the enhanced kinked node, $v_{tr}$ is the transverse speed of the kinked plasma, and $\Gamma$ is the bulk Lorentz factor of the jet. Here we characterize the QPOs in flux and polarization via the following analytical formulae.

We assume that the jet is axisymmetric and moving in the $z$-axis, pervaded by a helical magnetic field. By nature, kink instabilities are not axisymmetric physical processes, but we find that this approximation can greatly simplify the calculations while still capture the key QPO features of the kink-driven QPOs. A follow-up numerical simulation paper will show how this formula can match the key radiation patterns derived from MHD simulations. The poloidal magnetic field component, which is along the jet propagation direction, is called $B_z$. The toroidal component that is perpendicular to the jet direction is called $B_{\phi}$. In the linear stage of kink instabilities, the helical magnetic field will be twisted into a spiral shape \citep{haocheng_polarization_signatures}. But it is still a reasonable approximation to consider $B_z=a(r)B_0$ and $B_{\phi}=b(r)B_0$ as two orthogonal components of the magnetic field, where $a(r)$ and $b(r)$ describe the dependence of the magnetic field components on radius {\bf and $B_0$ is the average magnetic field strength}, although the definition of $B_{\phi}$ should be generalized to be the magnetic component perpendicular to the jet direction instead of just the toroidal magnetic field. It is well known that the blazar is observed very close to the line of sight (within $1/\Gamma$ from the jet direction), which corresponds to an angle $\theta$ between 0 and $90^{\circ}$ in the comoving frame of the jet. Since the magnetic field component parallel to the line of sight has minimal contribution to the observed synchrotron emission, we define the magnetic field component projected onto the comoving plane of sky that is perpendicular to the jet direction as $B_{\perp}$, and the parallel component $B_{\parallel}$. Then we have
\begin{equation}
\begin{aligned}
B_{\perp} & =-B_{\phi}\sin \phi \\
          & =-b(r)B_0\sin \phi \\
B_{\parallel} & = -B_{\phi}\cos \phi \cos \theta +B_z \sin \theta \\
              & = -b(r)B_0\cos \phi \cos \theta + a(r)B_0 \sin \theta
\end{aligned} ~~.
\end{equation}
We can find the synchrotron emission as
\begin{equation}
\begin{aligned}
I & \propto \int dr \, \int d\phi \, n(r) (B_{\perp}^2+B_{\parallel}^2) \\ 
  & \propto \pi B_0^2 \int dr \, n(r) (b^2(r)+b^2(r)\cos^2\theta+2a^2(r)\sin^2\theta)
\end{aligned} ~~,
\end{equation}
where $n(r)$ is the nonthermal electron density that makes the optical synchrotron emission in the observer's frame. We can similarly find the Stokes Q of the synchrotron emission
\begin{equation}
\begin{aligned}
Q & \propto \int dr \, \int d\phi \, \Pi(r) n(r) (B_{\perp}^2+B_{\parallel}^2) \cos 2\arctan{\frac{B_{\perp}}{B_{\parallel}}} \\
  & \propto \pi B_0^2 \int dr \, \Pi(r) n(r) (2a^2(r)\sin^2\theta - b^2(r)\sin^2\theta)
\end{aligned} ~~,
\end{equation}
where $\Pi(r)$ is the corresponding maximal synchrotron polarization degree. Here we take the convention that the polarization angle is 0 if the EVPA is perpendicular to the jet. Stokes $U$ is 0 given the axisymmetric assumption, and Stokes $V$ is 0 since isotropic synchrotron emission is generally linearly polarized. If we further assume that the nonthermal electrons follow the same power-law spectral index everywhere in the kinked jet, then $\Pi(r)=\Pi_0$ is a constant. Now we define $P_z\propto \pi B_0^2\int dr \, n(r)a^2(r)$ and $P_{\phi}\propto \pi B_0^2\int dr \, n(r)b^2(r)$, then we have the general expressions for synchrotron Stokes parameters in an axisymmetric jet as
\begin{equation}
\begin{aligned}
I & = P_{\phi}(1+\cos^2\theta)+2P_z\sin^2\theta \\
Q & = \Pi_0 (2P_z\sin^2\theta -P_\phi\sin^2\theta)
\end{aligned} ~~.
\end{equation}
If the blazar flux variability completely originates from nonthermal electron injection by kink instabilities, then $P_z$ and $P_{\phi}$ can be expressed as a steady component and a variable component. We assume that the steady components are $P_{z_0}=Z_0P_0$ and $P_{\phi_0}=P_0$, where $Z_0$ describes whether synchrotron emission from poloidal or toroidal magnetic component is stronger. Since the injection is preferentially in the central spine where the poloidal component is stronger, we assume that the synchrotron emission due to the quasi-periodic injection is completely in the poloidal component, $P_z(t)=F(t/T_1)P_0$, where $F(t/T_1)$ is a periodic function. Then we have
\begin{equation}
\begin{aligned}
I & = P_0(1+\cos^2\theta)+2(Z_0P_0+F(t/T_1)P_0)\sin^2\theta \\
  & = A_1+B_1F(t/T_1) \\
Q & = \Pi_0(2(Z_0P_0+F(t/T_1)P_0)\sin^2\theta-P_0\sin^2\theta) \\
  & = \Pi_0C_1+\Pi_0B_1F(t/T_1)
\end{aligned} ~~,
\end{equation}
where $A_1=P_0(1+\cos^2\theta)+2Z_0P_0\sin^2\theta$ and $C_1=2Z_0P_0-P_0\sin^2\theta$ are steady in time, and $B_1=2P_0\sin^2\theta$ is related to the periodic parts.

Alternatively, if the kinked jet propagates through a standing shock, the shock front can compress the toroidal magnetic field component and enhance the magnetic reconnection therein. This leads to quasi-periodic injection of nonthermal electrons following the spiral shape of the kink node, where the toroidal magnetic component is stronger. In this case, the period can be approximated by $T_2=\frac{L_{ki}}{v_{sh}\Gamma}$, where $L_{ki}$ is the length of the segment of the jet with strong kink instabilities, and $v_{sh}$ is the propagation speed of the shock in the comoving frame of the kinked jet. Following the above derivation, now the quasi-periodic injection is completely in the toroidal component, $P_\phi(t)=G(t/T_2)P_0$, where $G(t/T_2)$ is a periodic function, we have
\begin{equation}
\begin{aligned}
I & = (P_0+G(t/T_2)P_0)(1+\cos^2\theta)+2Z_0P_0\sin^2\theta \\
  & = A_2+B_2G(t/T_2) \\
Q & = \Pi_0(2Z_0P_0\sin^2\theta-(P_0+G(t/T_2)P_0)\sin^2\theta) \\
  & = \Pi_0C_2-\Pi_0B_2G(t/T_2)
\end{aligned} ~~,
\end{equation}
where $A_2=P_0(1+\cos^2\theta)+2Z_0P_0\sin^2\theta$ and $C_2=2Z_0P_0\sin^2\theta-P_0\sin^2\theta$ are steady in time, and $B_2=P_0\sin^2\theta$. This scenario will also be tested in the follow-up numerical simulation paper.

We can see that in either scenario the Stokes $I$ and $Q$ have strictly in-phase or anti-phase periodic patterns, although the amplitude and mean value are different. We note that the above formulae do not include turbulence or non-axisymmetric magnetic reconnection in the kink node, which can contaminate the periodic patterns. Additionally, turbulence does not contribute to Stokes $I$ and $Q$ equally as shown in previous simulations \citep{Zhang2023}. Nonetheless, as long as the nonthermal electron injection is dominated by kink instabilities, we should expect in-phase or anti-phase correlated QPOs in Stokes $I$ and $Q$ for kink-driven transient QPOs.

Under the leptonic scenario, the nonthermal electrons that produce the optical synchrotron emission can upscatter photons to \textit{Fermi}-LAT bands. The target photons can be the electron synchrotron itself or external radiation fields. Thus the $\gamma$-ray flux is given by
\begin{equation}
\begin{aligned}
F_\gamma & \propto \int dr \, \int d\phi \, n(\gamma)u_{ph} (r)
\end{aligned} ~~,
\end{equation}
where $u_{ph} (r)$ is the target photon energy density. This photon field can vary in time, which can complicate the temporal patterns. But as long as the flux variability is dominated by the quasi-periodic injection of $n(r)$, the $\gamma$-ray flux can be similarly expressed as a steady component plus a periodic component, similar to the above Stokes $I$. Given that in most blazar observations, \textit{Fermi}-LAT has continuous monitoring while optical data often has large gaps, we can treat the \textit{Fermi}-LAT data as the Stokes $I$ to look for in-phase/anti-phase correlated QPOs with the Stokes $Q$ of optical synchrotron emission.

We emphasize three observational features of the kink-driven QPOs that are not explicitly shown in the above analytical formulae and can affect the analyses of observational data. Firstly, the quasi-periodic injection leading to QPOs in Stokes $I$ and $Q$ has an implicit assumption that the injected electrons are mostly cooled within one cycle, i.e., the cooling time scale of these electrons should be much shorter than the period. For typical blazar parameters, this assumption means that the QPOs are only present near or beyond the cooling turnover of the spectrum. For low- and intermediate-synchrotron-peaked blazars, this corresponds to the neighborhood of the optical band (or \textit{Fermi}-LAT bands in the Compton scattering spectral component). For the high-synchrotron-peaked blazars, this corresponds to the X-ray and TeV bands. Secondly, the electron acceleration is through magnetic reconnection and turbulence mechanisms in kinked jet. Reconnection is known to make very strong and fast flares from numerical simulations \citep{Giannios2009,Zhang2022,Zhang2024}. Therefore, it is possible that different cycles of the kink-driven variability have different flare amplitudes and flux levels. This is beyond the traditional definition of QPOs, and we do not have a well established method to identify such temporal patterns. Hence, we restrict our search to the usual QPOs with similar flux levels and flare amplitudes, thus we will filter out observational epochs with high flux and strong variability. This does not mean that these epochs cannot result from kinked jet. Finally, we show that polarization signatures, though showing correlated QPOs with the flux, do not necessarily have the same period or the in-phase or anti-phase correlation. We use the QPOs driven by one or a part of the kink node as an example. It is clear from the expression for Stokes $Q$ that $C_1$ can be negative. Hence we have three possibilities here: 1. $C_1$ is positive; 2. $C_1$ is negative and the maximum of $Q$ is still negative, i.e., the maximum of $B_1F(t/T_1)$ is smaller than the absolute value of $C_1$; 3. $C_1$ is negative but the maximum of $B_1F(t/T_1)$ is larger than the absolute value of $C_1$, so $Q$ can change sign in time. In the former two cases, the in-phase or anti-phase correlation between Stokes $I$ and $Q$ directly result in in-phase or anti-phase QPOs in flux and polarization degree. The simulation in \citet{kink_qpos} falls into this category. In the latter case, however, the polarization degree can drop to zero then increase again when $Q$ changes sign. In this case, one cycle in flux may be accompanied by two bumps in polarization degree, and the polarization angle rotates two $90^{\circ}$, each corresponds to one bump in the polarization degree. The directions of two $90^{\circ}$ rotations depend on the non-axisymmetric and turbulent component of the kink instabilities, which cannot be modeled analytically. But they have only two options, either in the same direction to complete a $180^{\circ}$ angle swing, or in opposite directions. Additionally, whether the two rotations are in the same direction can vary cycle by cycle, and two neighboring cycles can have different rotation directions as well: they all depend on the temporal evolution of Stokes $U$. Therefore, the polarization angle evolution may show $180^{\circ}$ swing in one direction and then in the opposite direction for two neighboring cycles, as shown in \citet{kink_bllacertae}, or make multiple $180^{\circ}$ swings in the same direction. Consequently, it is best to use Stokes $I$ and $Q$ to find in-phase or anti-phase correlated QPOs to identify kink instabilities in blazars. Also, Stokes $I$ and $Q$ are both LCs, except that $Q$ can have negative values, which does not affect the search for QPOs. In the following tests, we will treat Stokes $I$ and $Q$ as LCs.

\section{Testing QPOs for Kink Events} \label{sec:test}
This section delves into two aspects of the search for transient QPOs in the context of kink events. Specifically, the first subsection presents a test to evaluate the detection of independent QPOs, evaluating whether they can be artifacts originating from noise. The second subsection focuses on determining the minimum number of oscillation cycles required to obtain a significant QPO, defined as $\geq$3$\sigma$. 

\subsection{Testing the Fake Detection of a Kink Event} \label{sec:fake_detection_test}
This study assesses the likelihood that two independent and random LCs exhibit statistical evidence to be associated with a kink event. To quantify this probability, we employ Monte Carlo simulations, creating an ensemble of 1,000,000 pairs of synthetic LCs using the method \citet{timmer_koenig_1995}, replicating the noise characteristics (red noise) and sampling patterns of the original observations, that is, 3-day binned (see $\S$\ref{sec:fermidata}), and with a duration of 2 years. 

For each artificial LC, we perform a periodicity search and record the maximum peak value in the resulting periodogram. This procedure is repeated for each synthetic LC to build a distribution of maximum peak values expected from random noise. We then use this distribution to derive confidence levels by identifying the peak values that correspond to specific significance levels (e.g., 1$\sigma$, 2$\sigma$, etc.). The significance of a given period in the original LC is estimated by comparing its peak value to these thresholds, allowing us to assign significance to the detected signal.

\paragraph{Testing Independent QPOs}
Initially, a test evaluates the detection of independent and compatible QPOs in a pair of LCs. For each pair of independent synthetic LCs, we perform a Lomb-Scargle periodogram \citep[LSP,][]{lomb_1976, scargle_1982} analysis to identify dominant frequencies and record instances where these frequencies coincide with a range of 10\%. This range is chosen according to the typical uncertainties in periods that are obtained in our study (see $\S$\ref{sec:results}). Additionally, we search for periods that ranged from a week to a maximum period of five months. The upper limit of five months can be estimated via the period formulae with typical blazar parameters. The one-week lower limit is due to the 3-day bin size of the data. By analyzing the distribution of these coincidences across many trials, we can establish a baseline probability for random period matching. This statistical framework allows us to determine whether the observed QPOs in the original LCs are likely the result of intrinsic periodic signals or mere chance. 

\begin{figure}
	\centering
	\includegraphics[scale=0.40]{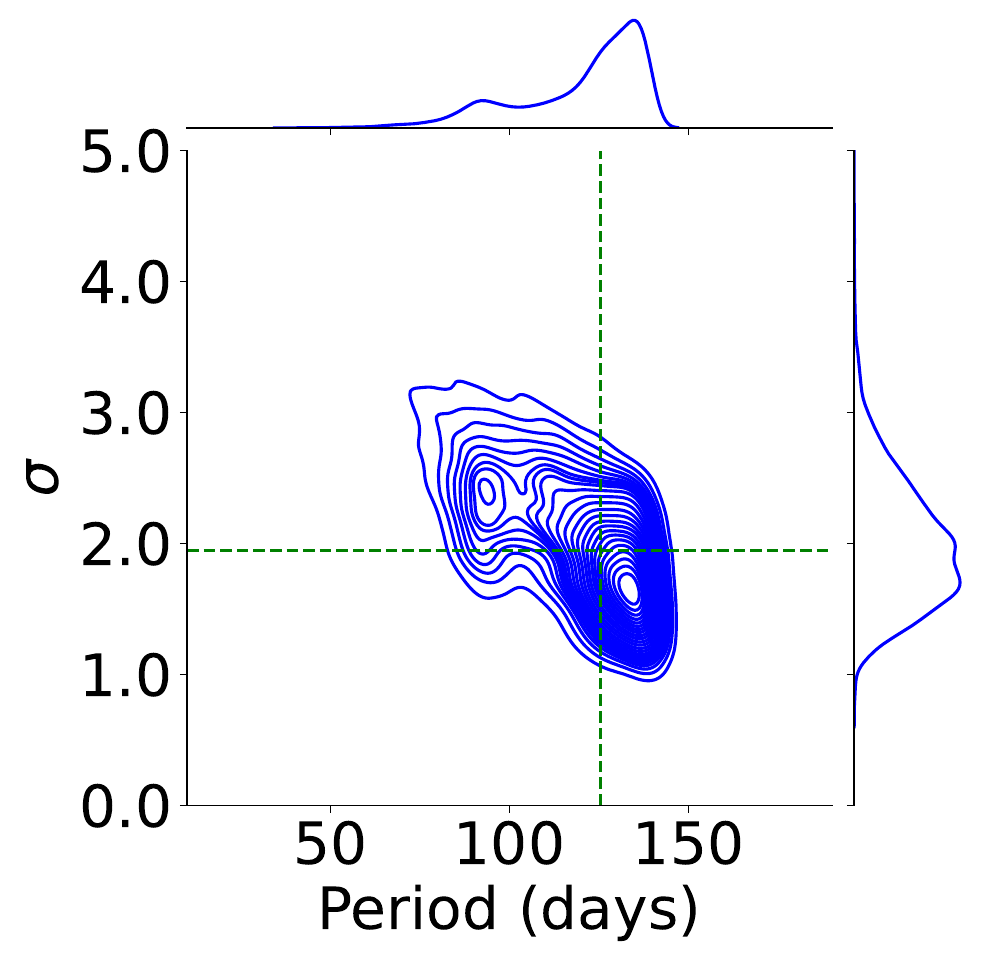}
		\caption{Distribution of the periods with a significance obtained by simulating with independent synthetic LCs with a compatible period. The median of significance and the periods (denoted by the dotted green lines) are 1.9$\sigma$ and 125 days, respectively.}
	\label{fig:test_1}
\end{figure}

As a result, we find that in 0.62\% of the cases, a compatible period is observed in a pair of synthetic LCs. This indicates a low probability of random coincidence, suggesting that such matches are relatively rare under the null hypothesis. The distribution of these periods and their associated significance is illustrated in Figure \ref{fig:test_1}. The median of the periods is 125 days, and the significance is 1.9$\sigma$, indicating that detected periods within this range and significance level could arise from stochastic processes. 

\begin{figure}
	\centering
	\includegraphics[scale=0.40]{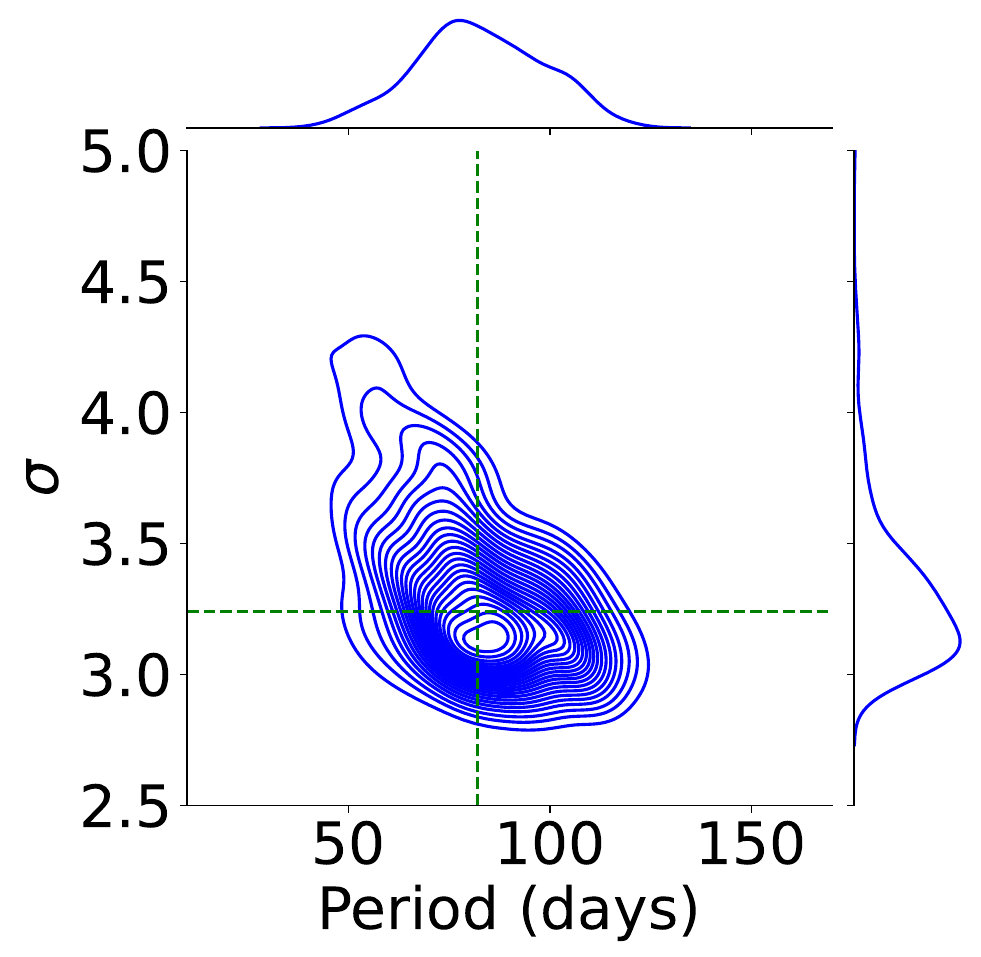}
		\caption{Distribution of the periods with an associated significance $\geq$3$\sigma$ of the sample  with compatible periods showing in Figure \ref{fig:test_1}. The median of significance and the periods (denoted by the dotted green lines) are 3.2$\sigma$ and 82 days, respectively.}
	\label{fig:test_1_sigma}
\end{figure}

\paragraph{Independent QPOs with 3$\sigma$}
For the 0.62\% of LC pairs with compatible periods, we further refine our selection by focusing on pairs where at least one LC has a period with a significance of $\geq$3$\sigma$. This threshold, as defined in $\S$\ref{sec:finer_analysis}, is set to identify periods that are statistically significant. As a result of this constraint, we find that 0.04\% of the LC pairs include at least one period with a $\geq$3$\sigma$ significance. The distribution of these periods and their corresponding significance levels is shown in Figure \ref{fig:test_1_sigma}. The median period for these significant pairs is 82 days, with a median significance of 3.2$\sigma$, indicating that only a small fraction of the periods meet our threshold for significance. This highlights that while compatible periods are detected, only a few are considered robust based on our criteria. 

\paragraph{Correlation of the Independent QPOs with 3$\sigma$}
Finally, for the pairs of LCs where at least one period has a significance of $\geq$3$\sigma$, we estimate the cross-correlation between the two LCs. We apply the criteria that the time lag between the two LCs should be approximately 0 days or close to half the period for in-phase or anti-phase correlation as described in the previous section. As a result of this analysis, $3\!\times\!10^{-4}$\% of the total pairs meet our methodology. This shows that the in-phase/anti-phase correlated Stokes $I$ and $Q$ criteria for kink-driven transient QPOs is highly unlikely due to red noise.

\begin{figure*}
	\centering
        \includegraphics[scale=0.40]{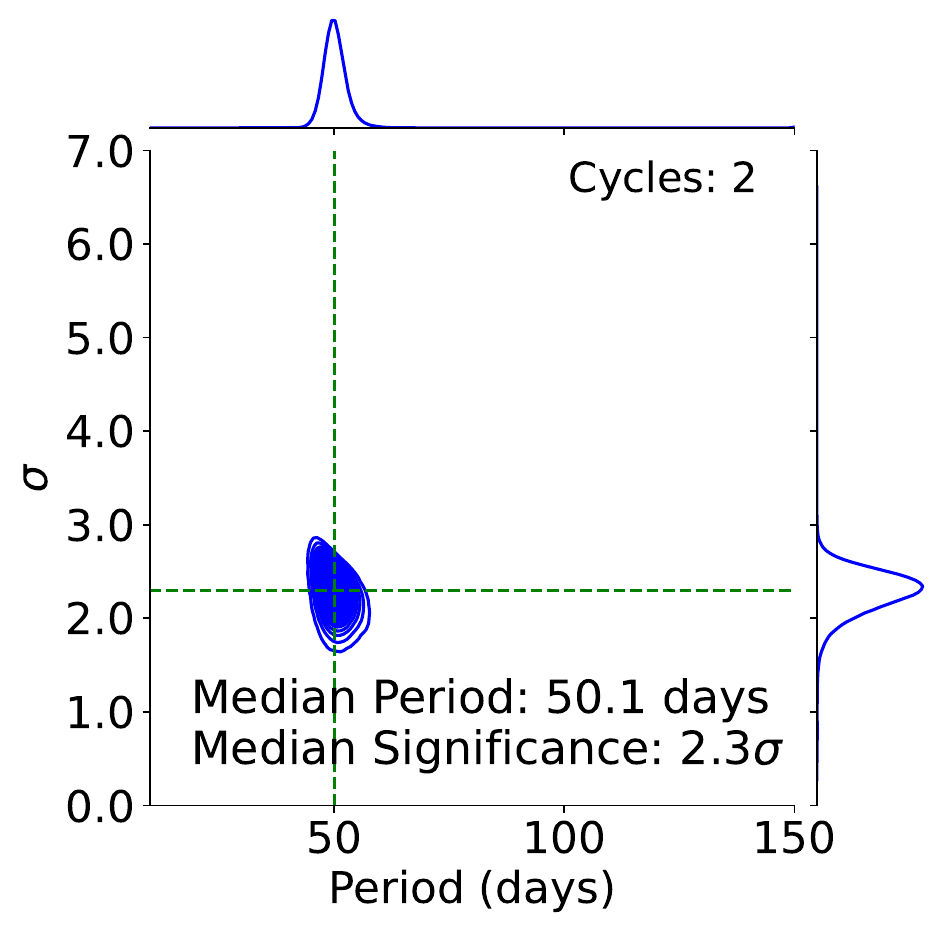}
	\includegraphics[scale=0.40]{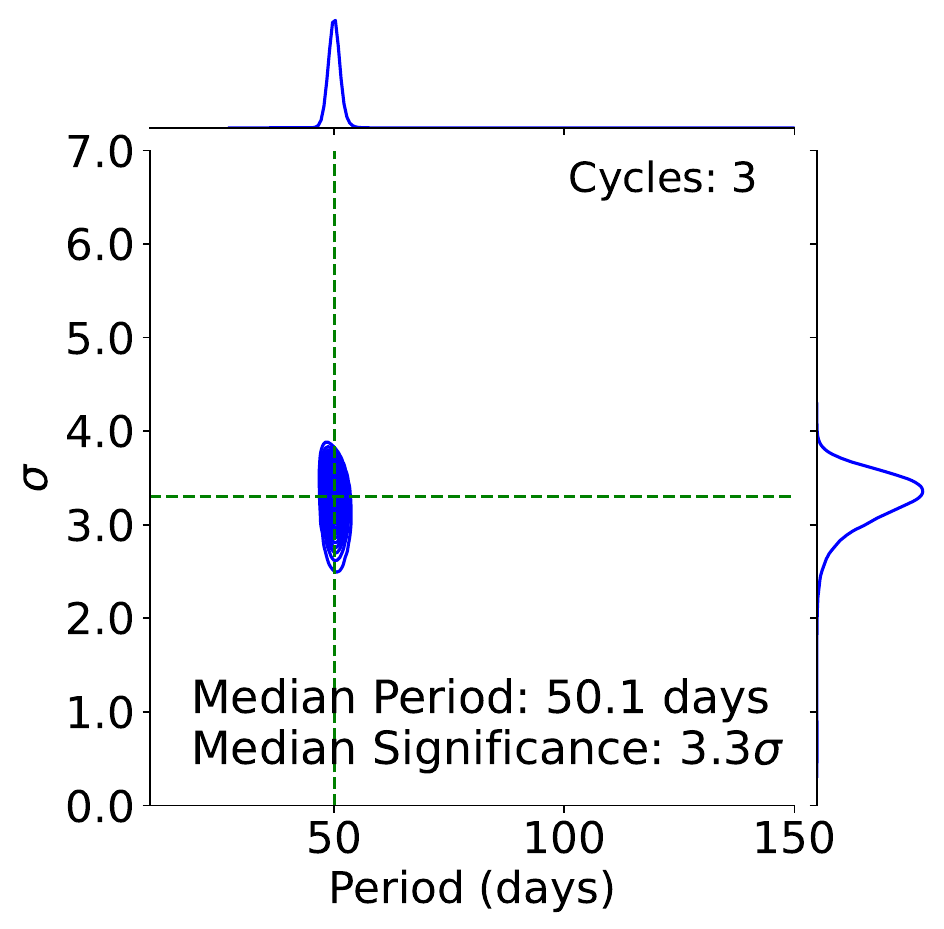}
        \includegraphics[scale=0.40]{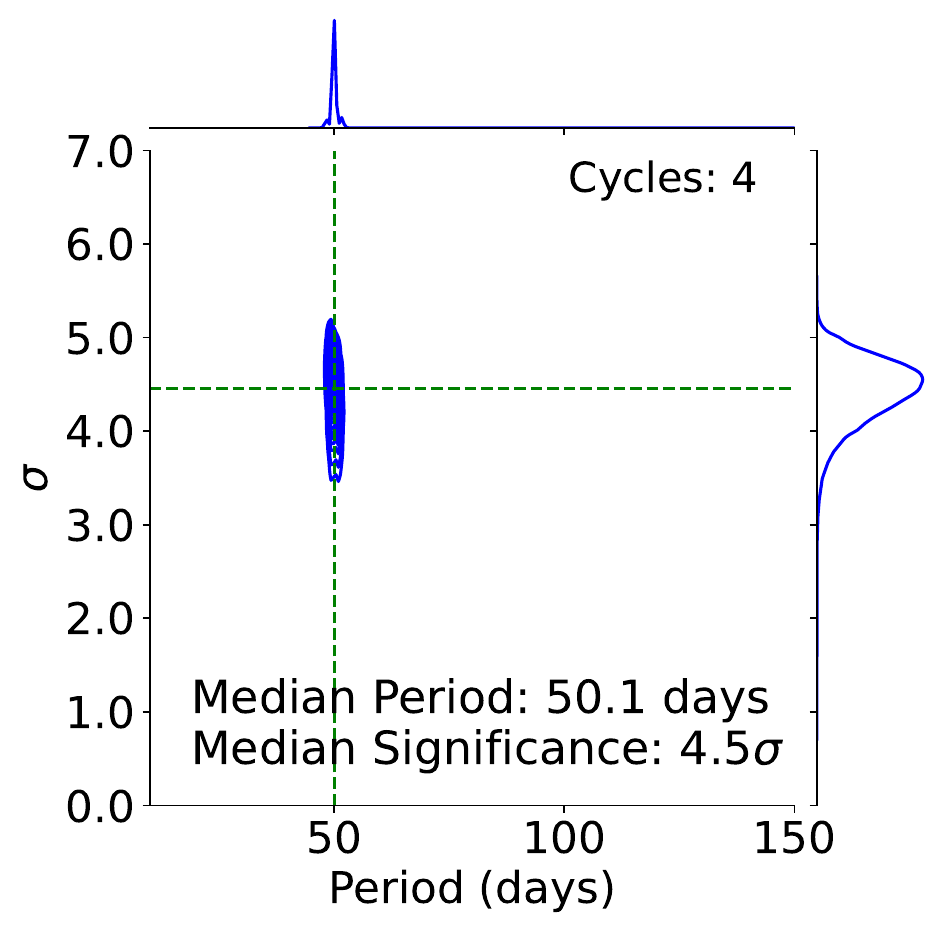}
        \includegraphics[scale=0.40]{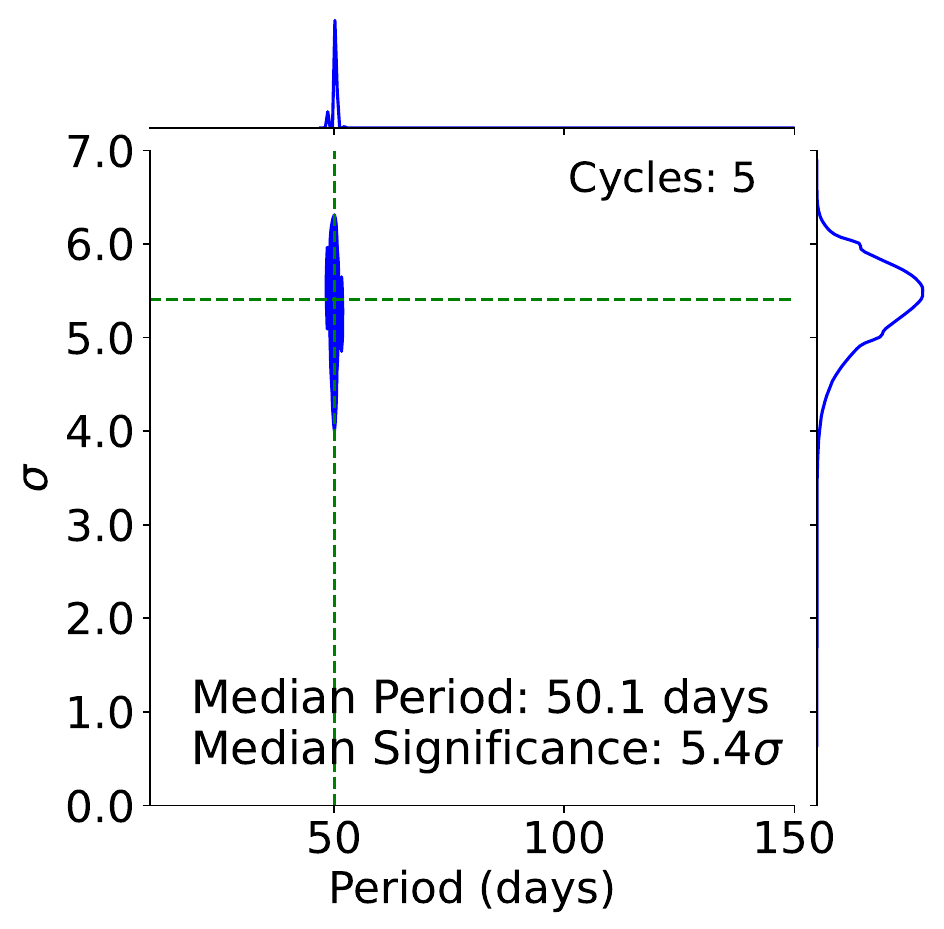}
		\caption{Distributions of the periods and the associated significance as a function of the number of cycles. The increase in significance is clearly associated with more cycles in the signal. As the number of cycles increases, the periods become more consistent, reducing the standard deviation of the period distribution.}
	\label{fig:num_cycles}
\end{figure*}

\subsection{Number of Cycles for a Significant QPO} 
This study evaluates the number of cycles required for a QPO to be considered significant. The purpose of this test is to establish criteria for selecting QPOs that have a minimum number of cycles, ensuring they are suitable for analysis in real data.

A QPO becomes statistically meaningful when the signal-to-noise ratio is sufficiently high, and this typically depends on the number of observed cycles. More cycles generally lead to higher confidence in the periodicity, as the periodic signal can be more reliably distinguished from the noise \citep{vaughan_criticism}. The required number of cycles for a QPO to be deemed significant can vary depending on the signal's strength and the noise's characteristics. 

To determine this number of cycles, we generate synthetic LCs with known red noise properties and introduce QPO signals of varying cycle numbers to determine the detection probability. Again, these synthetic LCs are generated using the method \citet{timmer_koenig_1995}. Gaussian noise (white noise) with a distribution of $N(0, 7.5\!\times\!10^{-6})$ is added to introduce stochastic variability to the sinusoidal signal points. This standard deviation for this noise is based on the highest standard deviations observed in the segments of LCs analyzed in our sample. We then estimate the significance of these QPO signals. This process is repeated 1,000,000 times to obtain a distribution of the periods and their associated significance levels. The significance is estimated by generating 1,000,000 red-noise LCs using the method \citet{timmer_koenig_1995}, and applying the same procedure than $\S$\ref{sec:fake_detection_test}.

Figure \ref{fig:num_cycles} shows the evaluation results of the number of cycles for a signal with a period of 50 days, which approximates the values of the periods presented in $\S$\ref{sec:results}. This figure illustrates the relationship between the number of cycles and the significance of the QPO detection. The period remains consistent with the original QPO across all cases, regardless of the number of cycles. As expected, the significance of the QPO improves with an increasing number of cycles, indicating that more cycles lead to a higher confidence in the detection of the periodic signal. The significance level of $\geq$3$\sigma$ is achieved with at least 3 cycles. 

However, it is important to consider some limitations of this experiment. First, the synthetic LCs, while replicating the statistical properties of the observed data, may not fully capture all the complexities present in real observations. Factors such as observational gaps and non-stationary noise components such as flares might affect the actual significance of detected QPOs in practice. We also assume that the period remains stable over time, but in real-world scenarios, QPO periods can drift or vary due to dynamic processes in the source, potentially complicating the detection and significance evaluation. Furthermore, we simulate a sinusoidal signal, which may not accurately reproduce the structure of the observed QPOs. All these factors must be considered when drawing conclusions from this experiment. It is clear, however, that the significance of a QPO improves with the number of cycles. 

Finally, these results are particularly relevant for interpreting transient QPOs, where the number of cycles is limited by the duration of the physical mechanism producing such QPOs. Understanding how significance scales with the number of cycles can help better evaluate the reliability of transient QPO detections and, thus, the limitation of the potential significance associated with such QPOs.

\section{Blazar Sample} \label{sec:fermidata}
We have selected 9 of the most variable blazars included in the 4FGL-DR3 catalog compiled by \textit{Fermi}-LAT \citep[][]{4fgl_dr3}. These blazars are chosen based on their significant variability, making them particularly suitable subjects for our search for QPOs in the timescale range associated with the kink events. In the context of 4FGL-DR3, a blazar is variable when the variability index is $\geq$24.725 \citep[as indicated by][]{4fgl_dr3}. Details regarding each of these blazars are listed in Table \ref{tab:candidates_list}.

\subsection{$\gamma$-ray Data} \label{sec:sample}

The LCs utilized in our analysis are sourced from the open-access \textit{Fermi} LAT Light Curve Repository \citep[][]{fermi_repository}\footnote{\url{https://fermi.gsfc.nasa.gov/ssc/data/access/lat/LightCurveRepository/about.html}}, which provides comprehensive data covering approximately 15 years of observations by the \textit{Fermi}-LAT, spanning from August 2008 to April 2023. For the purposes of this study, we have selected the 3-day binned LCs. This binning interval is strategically chosen to facilitate the search for QPOs spanning from a week to up to 5 months.

\begin{figure*}
	\centering
	\includegraphics[scale=0.21]{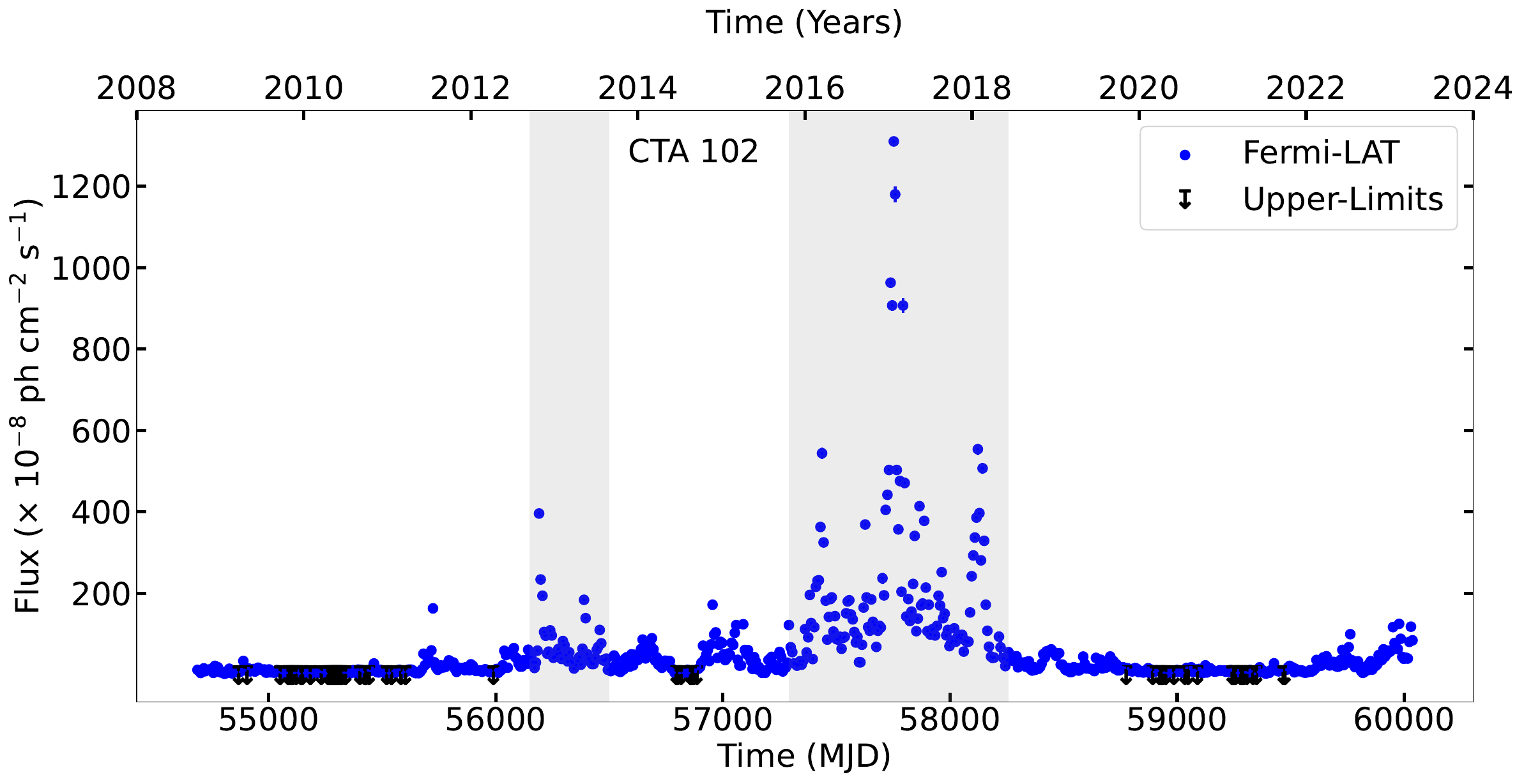}
        \includegraphics[scale=0.21]{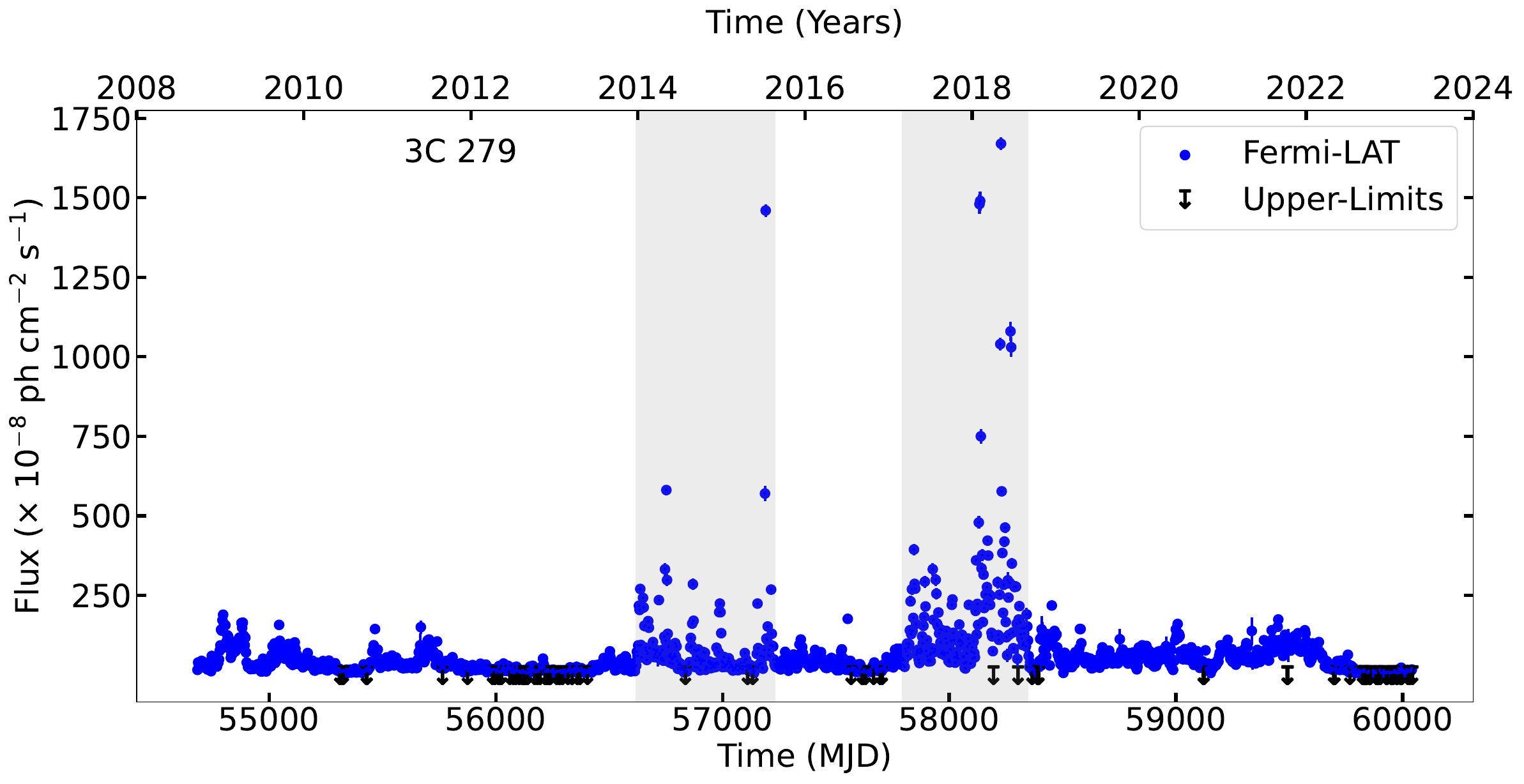}
		\caption{Examples of \textit{Fermi}-LAT Light curves of the blazars analyzed.The grey vertical lines denote the filtered sections. Right: CTA 102. Left: 3C 279.}
	\label{fig:lcs}
\end{figure*}

\subsection{Multiwavelength Data} \label{sec:wave_data}
For the MWL analysis, we employ archival data from different databases and observatories. We collect optical data, specifically in the V- and R- bands. The databases employed are Catalina Sky Survey \citep[][]{drake2009}\footnote{\url{http://nesssi.cacr.caltech.edu/DataRelease/}}, All-Sky Automated Survey for Supernovae \citep[][]{shappee2014,kochanek2017}\footnote{\url{http://www.astronomy.ohio-state.edu/asassn/index.shtml}},  American Association of Variable Star Observers\footnote{\url{http://https://www.aavso.org/data-download/}}, Small and Moderate Aperture Research Telescope System \citep[][]{bonning2012}\footnote{\url{http://www.astro.yale.edu/smarts/glast/home.php}}, and the Astronomy \& Steward Observatory\footnote{\url{http://james.as.arizona.edu/~psmith/Fermi/}} \citep{smith2009}, with optical V- and R-band observations. For the analysis, we combine all the data from different observatories. Different calibrations from different observatories can introduce offsets that affect the data. To evaluate and correct for the presence of such an effect, we compare all the simultaneous data from the different databases. If the simultaneous data show a systematic offset, the mean difference between each simultaneous pair of measurements is calculated. This estimation is used as the mean offset to correct the data from different observatories and ensure their compatibility.

We also use polarization data from different open-access and private databases. Specifically, we collect the polarization observations of our sample from archives Astronomy \& Steward Observatory (in the V-band), Robotic Polarimeter \citep[RoboPOL, data in R-band][]{robopol_blinov}\footnote{\url{https://robopol.physics.uoc.gr/}}, Multi-Optical-Band Polarization of Selected Blazars (MOBPOL, R-band)\footnote{\url{https://www.bu.edu/blazars/mobpol/mobpol.html}}, the Kanata Telescope \citep[R- and V- bands][]{kanata_polarization}, and the MAPCAT program carried out at the Calar Alto Observatory (CAHA) \citep{agudo2012}, Spain, with the 2.2-m telescope. The data in the R-band from the latter database was reduced with the IOP4 photo-polarimetric automatic pipeline \citep[see][]{escudero2024}.

\section{Methodology}\label{sec:methodology}
The search for signatures of kink events in the $\gamma$-ray LCs presents several challenges. The transient nature of kink-driven QPOs makes it difficult to perform a comprehensive QPO analysis on the entire LC, as short-duration QPOs can easily be obscured by the intrinsic noise within the data in the frequency domain. Moreover, the flux variability driven by kink can be contaminated by that from other nonthermal particle acceleration mechanisms and other parts of the jet. Consequently, there is a need for a specific methodology tailored to identify these transient QPOs. 

\subsection {Filtering High-Flux emissions}
The initial step in our methodology involves selectively filtering out segments of the $\gamma$-ray LCs that are characterized by high-flux emissions. High-flux flares in $\gamma$-ray blazar emission can significantly impact the search for kink events, making their identification and removal a crucial step in data analysis. These flares introduce substantial noise into the data, potentially masking underlying QPOs such as those linked to kink instabilities, complicating their detection \citep[][]{penil_flare_2025}. 

This is particularly necessary for the flaring states observed in blazar CTA 102 during approximately 2013 and again between 2016 and 2018, as well as the high-activity periods of 3C 279 in 2010-2011 and 2014-2016. Even if these very strong flares may contain kink-driven QPOs, they are beyond the applicability of our methodology since each flare can have very different flux levels. These high-flux states are depicted in Figure \ref{fig:lcs}. This approach allows us to concentrate on analyzing LC segments with more comparable flux levels. 

\subsection{Periodicity search methods}\label{sec:methods}
Following the initial filtering of high-flux segments in the LCs, we proceed to search for transitory QPOs. Our analysis segments the LCs into chunks of two years in duration. This timeframe is chosen to optimize the detection of QPOs, which ranges from a week to a maximum period of five months. 

The selection of two-year chunks ensures that multiple cycles of a potential QPO can be observed, even at the upper limit of the expected period. This is critical for addressing the concerns presented by \citet{vaughan_criticism} regarding the detection of spurious periodicities in LCs that contain only a few observable cycles. By ensuring that at least three cycles can be observed, we enhance the reliability of our periodicity analysis and reduce the likelihood of falsely identifying noise as true periodic signals \citep{vaughan_criticism}. 

To search for QPOs, we employ two methods: the Generalized Lomb-Scargle periodogram and the Singular Spectrum Analysis. The Generalized Lomb-Scargle periodogram \citep[GLSP,][]{lomb_gen} builds upon the traditional LSP by incorporating the measurement uncertainties of the data points into the analysis. By considering these uncertainties, the GLSP includes the influence of data error on the detection of periodic signals, thereby enhancing the reliability of the period estimates.

The Singular Spectrum Analysis \citep[SSA,][]{ssa_greco}, on the other hand, is a technique for decomposing a time series into its constituent components, separating the signal from the noise. SSA reconstructs the underlying deterministic components of a time series, effectively filtering out random fluctuations such as noise and flares. This capability makes SSA particularly useful for clarifying the oscillatory behavior in the original LCs, abstracting the impact of the inherent noise of the LC. Once SSA has isolated the oscillatory behavior from the stochastic components, we apply the LSP to this refined signal. This two-step approach allows us to accurately infer the periodic nature of the signal and determine the uncertainty associated with the identified period \citep{alba_ssa}. 

\subsection {Pre-selection QPOs}
\begin{figure}
	\centering
	\includegraphics[scale=0.41]{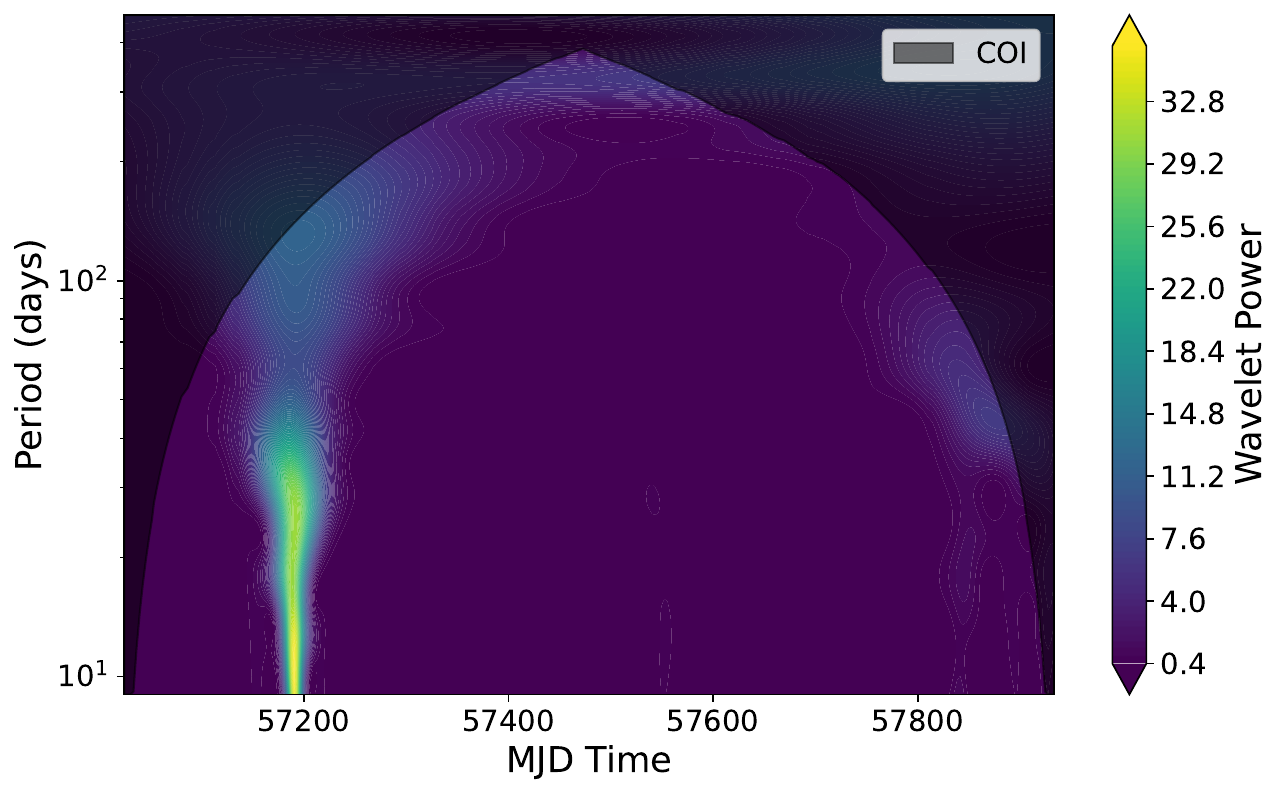}
		\caption{The CWT analysis of a selected LC chunk of 3C 279 is shown. The colormap represents the QPO information over time, with colors indicating the signal's power spectrum. This visualization helps identify variability patterns and highlights potential QPO presence. The shaded region, known as the cone of influence (COI), marks areas where potential modulations may be less reliable due to their proximity to the sampling interval or the signal's total duration. The figure suggests a potential QPO occurrence around MJD 57200.}
	\label{fig:3c279_cwt}
\end{figure}

Due to the transient nature of kink events, the associated QPOs are challenging to identify in the complete LC. Therefore, an initial search is necessary to detect the potential presence of such QPOs within specific segments of the LC. This selection is carried out using the Continuous Wavelet Transform \citep[CWT][]{wavelet_torrence}, which helps pinpoint the segments where QPOs could occur (see Figure \ref{fig:3c279_cwt}).

Once the relevant chunks are identified, further refinement is performed by focusing on the segments where QPOs are most observable. This targeted approach allows us to concentrate on the LC portions where QPOs are most pronounced, effectively excluding segments without QPOs. Such exclusion is crucial, as the presence of non-QPO data can weaken the statistical significance of our analysis and potentially obscure true periodicities. By isolating and analyzing only the relevant segments, we enhance the reliability of our findings and improve the likelihood of accurately identifying significant QPOs. 

However, this procedure may introduce bias due to ``cherry-picking,'' which occurs when data points that align with a pre-existing hypothesis are selectively emphasized. Such bias can lead to false conclusions about the presence of periodic signals. Although this bias is inherent and cannot be fully eliminated by subsequent analysis methods, we apply several multiple tests to ensure robust results. These steps include verifying the independence of detected QPOs, assessing their possible origin as stochastic noise, and requiring a minimum number of oscillation cycles for confident detection ($\S$ \ref{sec:test} and $\S$\ref{sec:finer_analysis}). We also perform the look-elsewhere effect in $\S$\ref{sec:global_signifcance}.

In this analysis stage, a total of 12 chunks were selected across all the blazars in our sample.

\subsection {Finer Analysis} \label{sec:finer_analysis}
QPO searches in astrophysical time-series data are significantly limited by noise, which often masks or mimics true periodic signals. This noise is associated with erratic brightness fluctuations characterized by steep power spectra, commonly referred to as red noise \citep[][]{rieger_2019}. This type of noise increases in power at lower frequencies and is notorious for leading to the detection of spurious periodicities \citep[][]{vaughan_criticism}. Additionally, the observational challenge is compounded by Poisson noise, which arises from the random nature of photon detection. This type of noise introduces variability into the LCs of blazars, particularly in $\gamma$ rays, adding a constant white noise component to the power spectrum.

The selected segments of LCs undergo a detailed analysis due to concerns that the method by \citet{timmer_koenig_1995} might overestimate significance \citep{jorge_2022}. To address this, we model the power spectral density (PSD) of these chunks using a power-law function of the form $A*f^{-\beta}+C$ where $A$ is the normalization, $\beta$ represents the spectral index, $f$ is frequency, and $C$ is the Poisson noise level. The parameters of this model are estimated using Maximum Likelihood estimation complemented by Markov Chain Monte Carlo (MCMC) analysis\footnote{We utilize the Python package emcee}. Following the PSD modeling, we apply the technique developed by \citet{emma_lc} to generate synthetic LCs that replicate both the PSD and the probability distribution function of the original data. For this purpose, we generate 1,000,000 synthetic LCs using the procedure outlined in \citet{connolly_code}. 
In this analysis, the local significance estimated for both methods is obtained as the fraction of simulated LCs in which the power at any frequency within the studied range exceeds the highest peak observed in the original LC.

In the final phase of analysis, we apply new selection criteria to these segments, retaining only those where the detected periods exhibit local significance $\geq$3$\sigma$ in either GLSP or SSA analyses. This threshold ensures consistency with the tests in $\S$\ref{sec:test}, allowing only significant QPOs to be considered in our results.

Finally, we apply the above analysis to the Stokes $Q$ in the same segment to check if it shows the same period. If so, we utilize the \textit{z}-transformed Discrete Correlation Function (\textit{z}-DCF) \citep[\textit{z}-DCF,][]{zdfc_alexander}, which is designed to analyze time lags in unevenly sampled datasets. This method enhances the traditional Discrete Correlation Function by addressing biases introduced by uneven time sampling in observations \citep[][]{zdfc_alexander}. The \textit{z}-DCF will allow us to measure time lags between the two Stokes parameters, identifying whether they are in-phase (lag$\approx$0), anti-phase (lag$\approx$$period/2$). 

\subsection {Look-Elsewhere Effect} \label{sec:global_signifcance}
In our periodicity analysis, there was no prior knowledge of the frequencies of the potential signal. In these conditions, it is statistically more rigorous to employ a ``global significance''. This significance accounts for the look-elsewhere effect, which is the ratio between the probability of observing the excess at some fixed value and the probability of observing it anywhere in the value range considered in the analysis \citep{gross_vitells_trial}. The ``global significance'' is obtained by applying a correction to the local significance of the periodicity in an LC at a specific value obtained for each method. This correction is approximated by
\begin{equation}\label{eq:trial}
p_{\mathrm{global}}=1-(1-p_{\mathrm{local}})^{N},
\end{equation}
where $N$ is the trial factor. In our study, we have to consider two potential issues. One is that we do not know the frequency for each source a priori, so we must search for the highest peak in each periodogram. The second issue is that we do not know a priori which sources exhibit periodic behavior, so we must also select them from the periodograms. Consequently, searching $P$ independent periods (frequencies) in each of the periodograms of $B$ blazars, the number of trials is 
\begin{equation}
\label{eq:p}
 N=P \times B.    
\end{equation}
 We do not consider the number of methods in the trials since we present all the results for all the methods equally in our tables for each blazar, avoiding picking the highest significant result according to a single method.

\begin{figure}
	\includegraphics[width=\columnwidth]{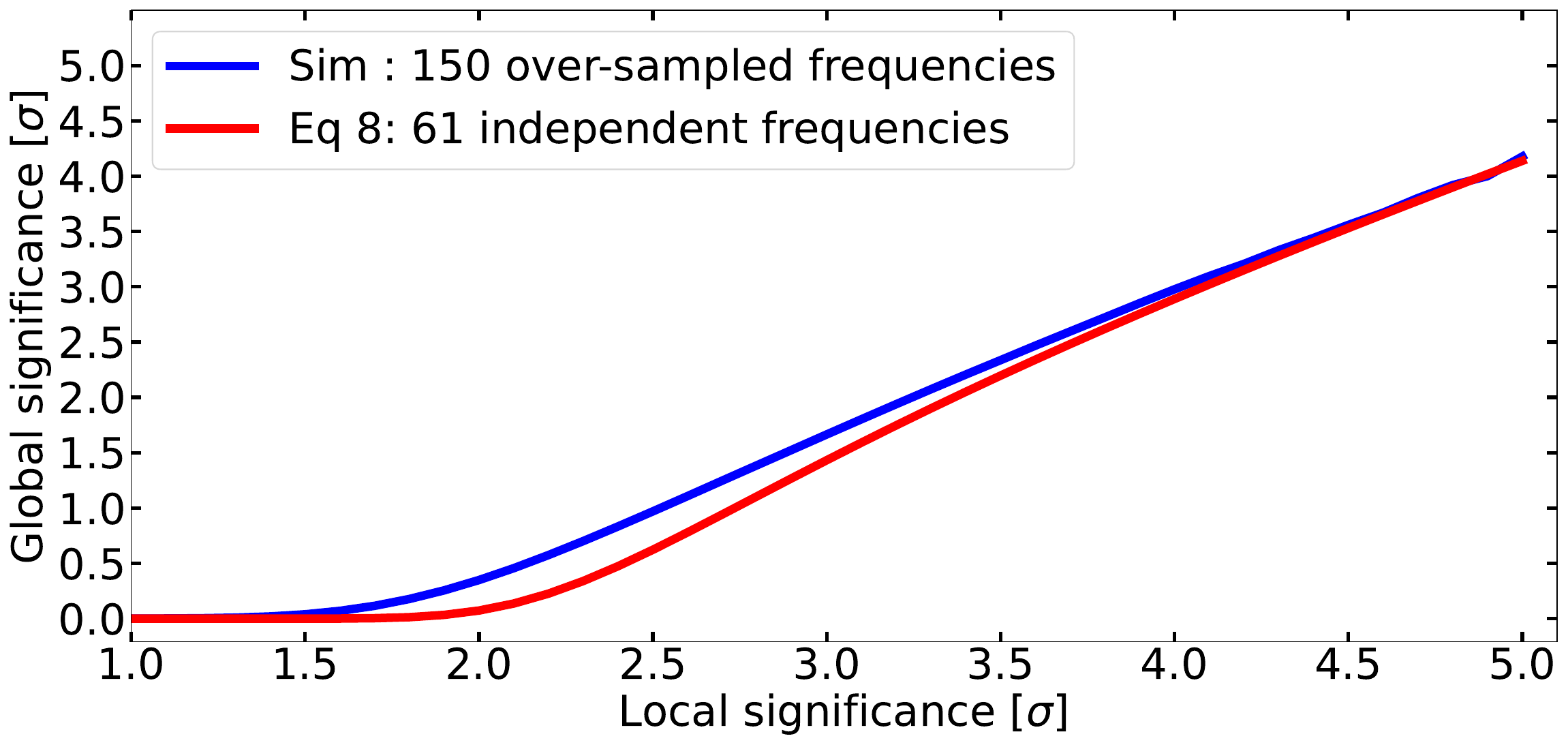}
	\caption{Trial-factor correction applied to the experimental relationship between local-global significance to estimate {\it P} in the periodicity analysis. {\it Eq. 8} represents the results obtained by applying Equation \ref{eq:trial} for a particular number of independent frequencies, reporting 61. The $P$ is chosen to adjust the local significance of $\approx4.0\sigma$, thereby improving the correction of the most significant period of Table \ref{tab:results}.} \label{fig:trials}
\end{figure}

In our periodograms, we incorporate 150 periods to strike a balance between computational efficiency and resolution. The estimation of $P$ is conducted through Monte Carlo simulations, specifically employing the algorithm described in \citep[][]{penil_2022}. By utilizing $10^{8}$ simulated LCs using the technique of \citet{timmer_koenig_1995}, we obtain the empirical relationship between local-global significance (as depicted by the blue line in Figure \ref{fig:trials}). To determine the most suitable $P$, we explore various values by applying Equation \ref{eq:trial} to best align with the experimental relationship of local-global significance. The selection of $P$ aims to correct the ``local significance" to be $\approx4.0\sigma$, according to the most significant periods of  Table \ref{tab:results}. 

Figure \ref{fig:trials} shows that $P=61$ provides the best results with the empirical relationship of local-global significance. Consequently, this choice leads to a trials factor of 61 ($9$$\times$$61$). Subsequently, according to this correction, the ``global significance'' of the periods presented in Table \ref{tab:candidates_list}:

\begin{itemize}
	\item $\approx$2.1$\sigma$ for a local significance of $4.0\sigma$ 
	\item $\approx$1.2$\sigma$ for local significance of $\approx$3.5$\sigma$
	\item $<$1$\sigma$ for a local significance $<$3.5$\sigma$.
\end{itemize}

\section{Results} \label{sec:results}
\begin{figure*}
	\centering
        \includegraphics[scale=0.21]{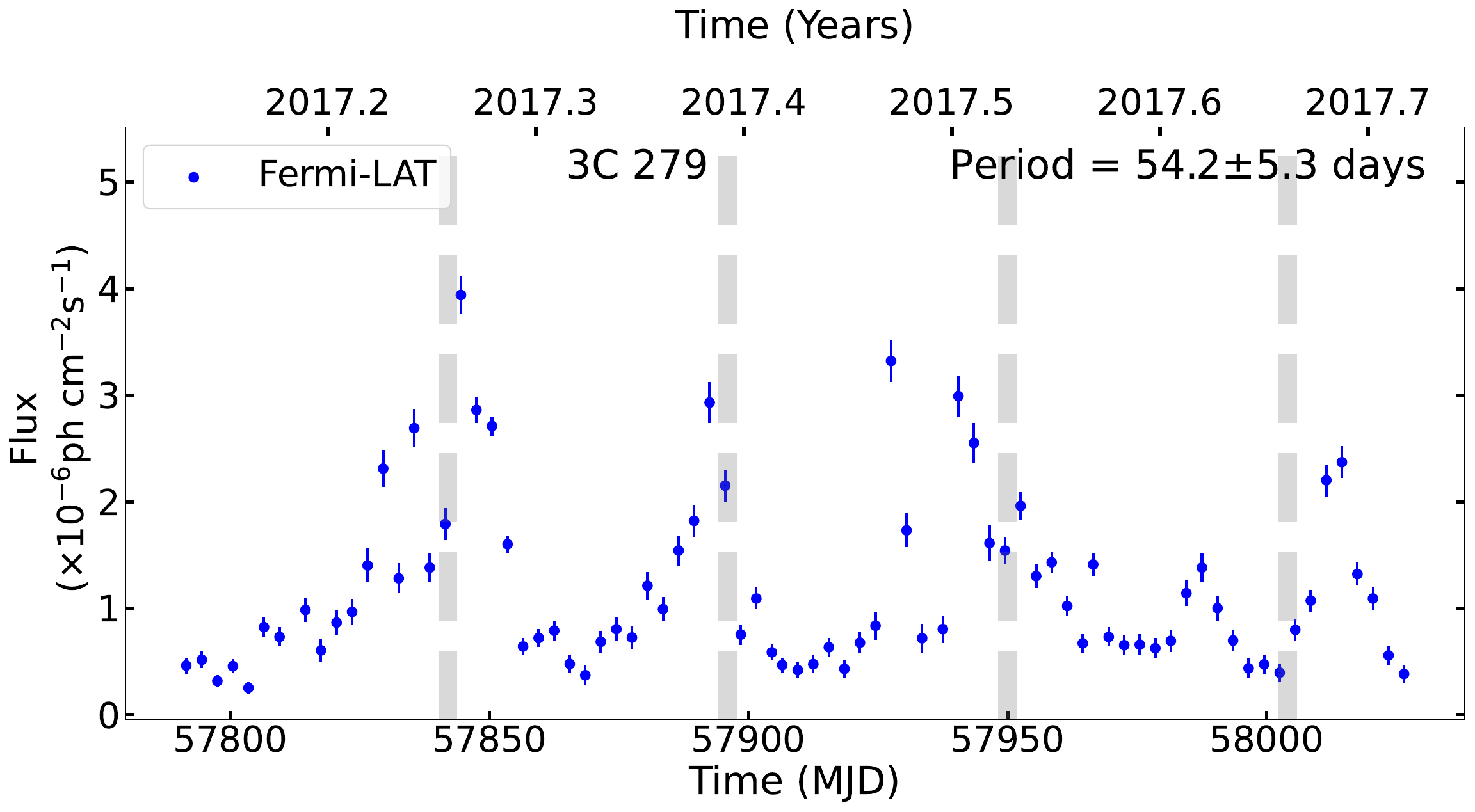}
	\includegraphics[scale=0.21]{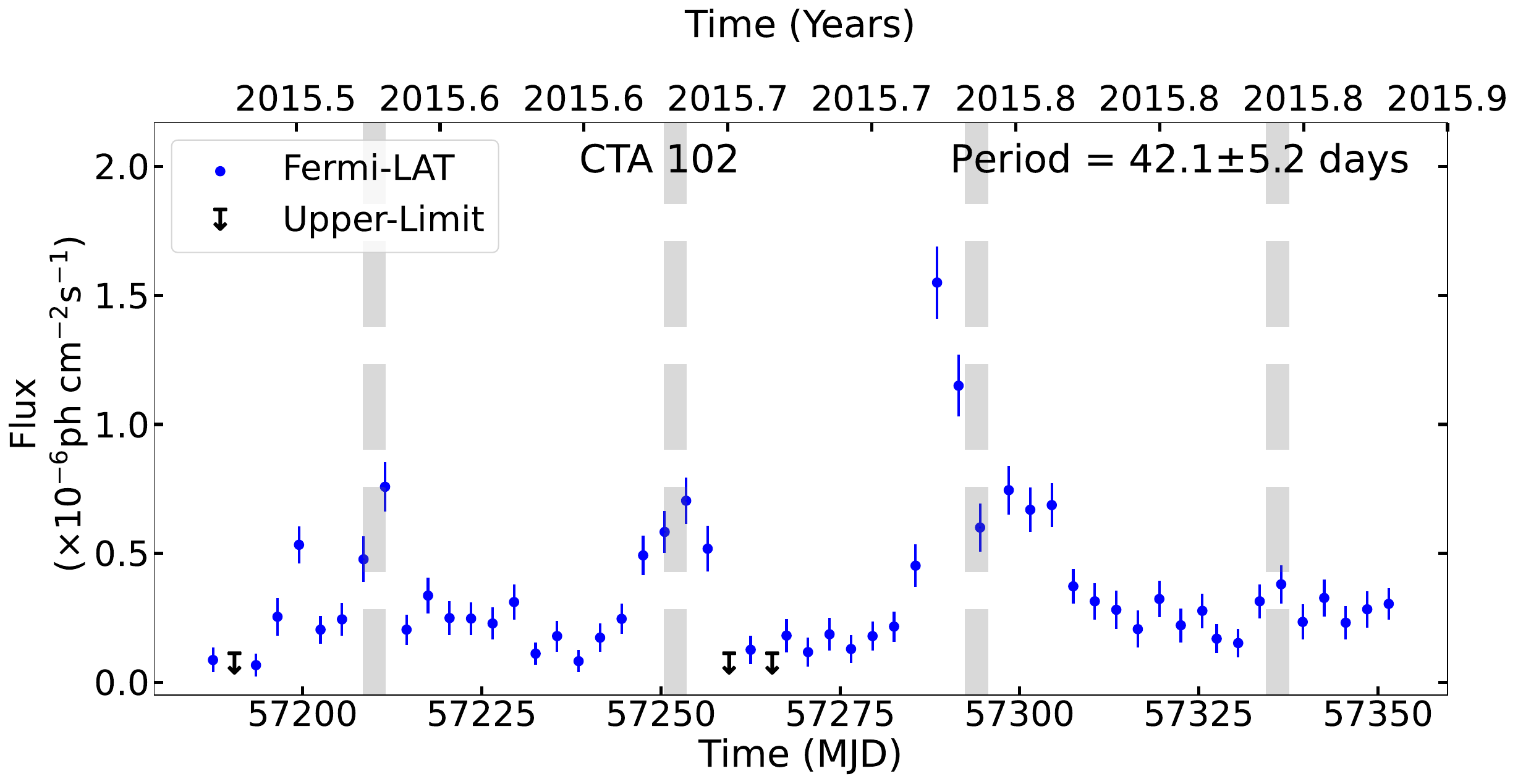}
		\caption{\textit{Left}: QPO of 3C 379, section 57790-58050. \textit{Right}: QPO of CTA 102. The gray vertical bars approximate high-flux periods suggested by the period inferred by the methodology. The width of the gray bars indicates the uncertainty in the periodic signal.}
	\label{fig:cta_qpo}
\end{figure*}

As a result, we find evidence of QPOs in the $\gamma$ rays in 4 of the blazars for a total of 6 segments, listed in Table \ref{tab:results}: 
\begin{itemize}
    \item For 3C 66A, we identify a segment 57790-58025 with a period of 52.1 days (2.2$\sigma$).
    \item For 3C 279, we identify 3 segments (57200-57650, 57790-58050, 58520-59060) with QPOs with 63.8 days (3.4$\sigma$), 54.2 days (2.8$\sigma$) and 86.7 days (3.2$\sigma$), respectively (see example in Figure \ref{fig:cta_qpo}).
    \item For CTA 102, we identify a segment 57170-57350 with a period of 45.0 days (2.7$\sigma$), shown in Figure \ref{fig:cta_qpo}. 
    \item For 3C 454.3, we identify a segment 58040-58220 with a period of 41.6 days (3.3$\sigma$).
\end{itemize}
 
No significant QPOs are observed in 3C 273, Mkn 421, 4C +21.35, Mkn 501, and PKS 1830$-$211.

From these QPOs detected in $\gamma$ rays, we evaluate the multi-wavelength emissions in order to check if similar QPOs are observed.

\subsection{Optical Analysis}
In the figures provided in Appendix \ref{sec:appendix}, the optical LCs present a limitation for conducting a periodicity analysis due to limited data availability. Only in the case of CTA 102 (Figure \ref{fig:parts_cta}) is it possible to search for QPOs. Consequently, we determine a period of 45.5$\pm$7.4 days (2.7$\sigma$) and 46.2$\pm$5.9 days (3.3$\sigma$) for GLSP and SSA, respectively. These values align with the period derived from $\gamma$ rays, which is 45 days.

\begin{figure*}
	\centering
	\includegraphics[scale=0.12]{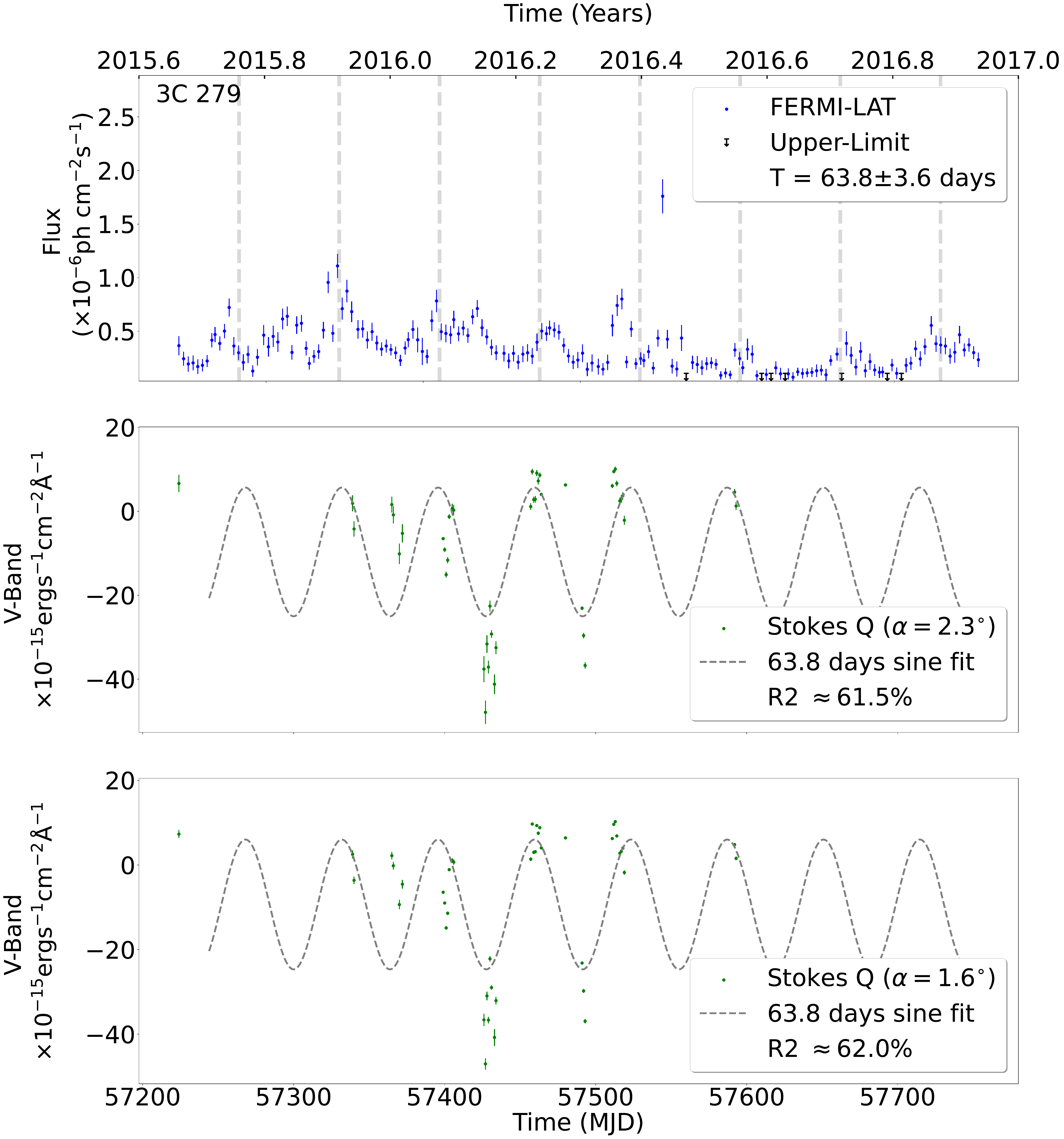}
		\caption{QPO of 3C 379 of section 57200-57650 with a period of 63.7 days. \textit{Top}: $\gamma$-ray LC. The gray vertical bars approximate high-flux periods suggested by the period inferred by the methodology. The width of the gray bars indicates the uncertainty in the periodic signal. \textit{Center}: LC of the Stokes $Q$ of the V-band, estimated by using the $\alpha=$1.6. \textit{Bottom}: LC of the Stokes $Q$, estimated by using the $\alpha=$2.3. The grey dotted line represents the sinusoidal reconstruction based on the period inferred from the $\gamma$-ray. R$^{2}$ calculated from the data, and the sinusoidal regression denotes a ``moderate'' fit \citep{hair_r2_2011}.}
	\label{fig:3c279_sine}
\end{figure*}

\begin{figure*}
	\centering
	\includegraphics[scale=0.12]{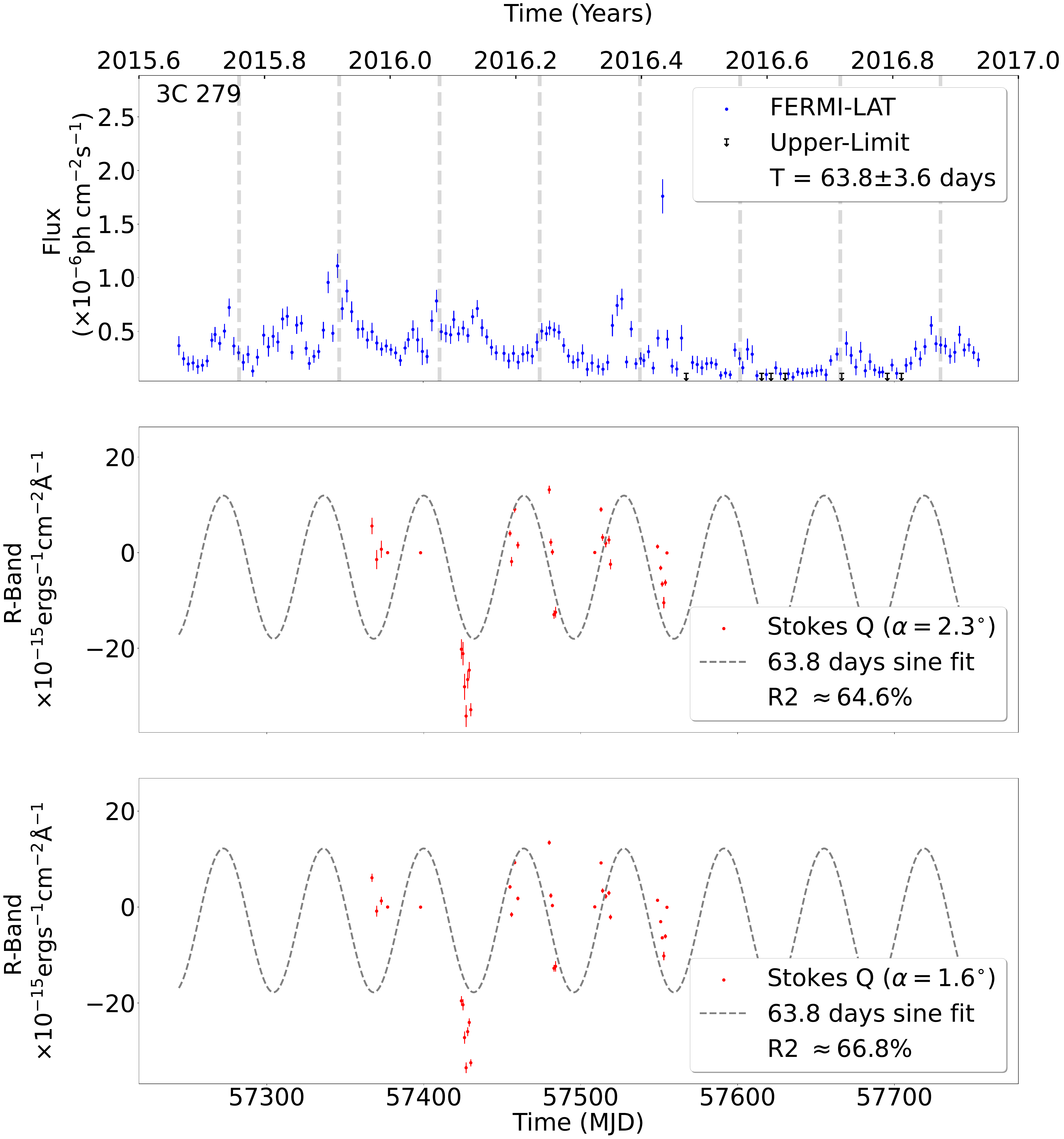}
		\caption{QPO of 3C 379 of section 57200-57650 with a period of 63.7 days. \textit{Top}: $\gamma$-ray LC. The gray vertical bars approximate high-flux periods suggested by the period inferred by the methodology. The width of the gray bars indicates the uncertainty in the periodic signal. \textit{Center}: LC of the Stokes $Q$ of the R-band, estimated by using the $\alpha=$1.6. \textit{Bottom}: LC of the Stokes $Q$, estimated by using the $\alpha=$2.3. The grey dotted line represents the sinusoidal reconstruction based on the period inferred from the $\gamma$-ray. R$^{2}$ calculated from the data, and the sinusoidal regression denotes a ``moderate'' fit \citep{hair_r2_2011}.}
	\label{fig:3c279_sine_rband}
\end{figure*}

\subsection{Stokes Parameters}

The presence of a kink event can be enhanced by analyzing the Stokes parameters, $I$ and $Q$. Specifically, the detection of a similar QPO as $\gamma$ rays supports the hypothesis that these periodicities might have originated from kink events within the jet structure \citep[][]{haocheng_polarization_signatures}. 

To illustrate this, we analyze the data of Figure \ref{fig:parts_3c279_1} of 3C 279, since the QPO is $\geq$3.5$\sigma$ in the $\gamma$ rays and this segment presents more available data than in other segments of the same object and the other objects. This analysis requires knowledge of the jet angle, $\alpha$, for the source 3C 279. According to \citet{Jorstad_angle_3c279}, the jet angle $\alpha$ has been estimated using two different methods, yielding values of 1.6$\pm$1.0 and 2.3$\pm$2.2 degrees, respectively. We calculate the Stokes parameter $Q$ for both jet angles, which is shown in Figure \ref{fig:3c279_sine}. Similar calculations are done for the R-Band, also shown in Figure \ref{fig:3c279_sine_rband}.

Due to limited data, however, a thorough analysis of any phase or anti-phase relationship between $I$ and $Q$ is not feasible at this time. Once again, the data quality is compromised by numerous gaps, which prevent a consistent and robust analysis. These interruptions make it impossible to conduct a feasible study to reliably detect the presence of a kink event. The lack of continuous data severely limits our ability to identify any patterns or features that might suggest such an occurrence. Consequently, any conclusions drawn regarding kink events remain speculative and warrant further investigation with more complete datasets in the future. As illustrated in Figure \ref{fig:parts_3c66a}, Figure \ref{fig:parts_3c279_1}, Figure \ref{fig:parts_3c279_2}, Figure \ref{fig:parts_3c279_3}, Figure \ref{fig:parts_cta}, and Figure \ref{fig:parts_3c454} the shortage of data limits our ability to conduct a comprehensive analysis. 

A compatible method can be used to evaluate whether the Stokes $Q$ shows evidence of a periodic pattern, especially when the available data does not support QPO analysis or phase/anti-phase relationship evaluation. Specifically, we perform a sinusoidal fitting using the period identified from the $\gamma$-ray LCs. To assess the quality of this fit, we use the R-squared (R$^2$) statistic, which measures the goodness of fit for regression models and ranges from 0 to 1. Higher R$^2$ values indicate a better fit. This approach provides a metric to infer whether the analyzed time series is plausibly described by a sinusoidal model with a specific period.

For the time segment from MJD 57200 to 57650 of 3C 279 (Figure \ref{fig:3c279_sine}), the R$^2$ value for the Stokes $Q$ in the V-band is approximately 60\%, indicating a ``moderate'' fit according to the criterion of \citep{hair_r2_2011}. Similarly, in the R-band, R$^2$ is approximately 65\%, also classified as a ``moderate'' fit. The presence of extensive gaps introduces significant uncertainty, making it difficult to draw a definitive conclusion regarding the existence of a QPO in the Stokes $Q$. Nevertheless, this result is valuable in supporting the potential kink origin of the observed QPO.

\section{Summary} \label{sec:summary}
This paper presents an investigation into the detection of kink events by searching for transient QPOs in blazars. We outlined the observational conditions required to examine that kink events are the physical origin of these QPOs. Our procedure for seeking evidence of kink events leverages both $\gamma$-ray and polarized data observations. Detecting the correlated QPOs in both flux and polarized flux is essential for confirming kink instabilities as the physical driver for such periodic behaviors.

In our study, we pinpointed 6 specific segments across 4 blazars --3C 66A, 3C 279, CTA 102, and 3C 454.3-- where significant QPOs were observed in their $\gamma$-ray emission. These QPOs exhibited periodicities ranging from 45 to 85 days. We also analyzed polarized data that corresponded to the temporal segments where these QPOs were initially detected. Unfortunately, the scarcity of comprehensive polarized data restricted our capability to conduct an extensive analysis of these QPOs. Despite these limitations, we identified one particular segment where the same QPO might appear in the associated polarized data. This observation provides tentative evidence supporting the hypothesis that this QPO may originate from kink instabilities in the blazar jets. To conclusively identify kink-driven transient QPOs in blazars, it is necessary that \textit{Fermi}-LAT can trigger comprehensive optical polarization monitoring when three cycles of QPO patterns are detected in $\gamma$-rays, an indication of significant QPOs based on our study, and then the optical polarization monitoring can establish $\ge$ 3 cycles of in-phase/anti-phase correlated Stokes Q of the same period with the $\gamma$-ray band.

\software{
    astroML \citep{astroml},
    astropy \citep{astropy_2013, astropy_2018, astropy_2022},
    emcee \citep {emcee}, 
    PyCWT \url{https://pypi.org/project/pycwt/},
    SciPy \citep {SciPy},
    Simulating light curves \citep{connolly_code},
    Singular Spectrum Analysis \url{https://www.kaggle.com/code/jdarcy/introducing-ssa-for-time-series-decomposition},
    \textit{z}-DCF, \citep{zdfc_alexander},    
}

\section{Acknowledgements}

The IAA-CSIC team acknowledges financial support from the Spanish “Ministerio de Ciencia e Innovación” (MCIN/AEI/ 10.13039/501100011033) through the Center of Excellence Severo Ochoa award for the Instituto de Astrofísica de Andalucía-CSIC (CEX2021-001131-S), and through grants PID2019-107847RB-C44 and PID2022-139117NB-C44. We also want to thank all the observatories from which we used data. We thank the Las Cumbres Observatory and its staff for their continuing support of the ASAS-SN project. ASAS-SN is supported by the Gordon and Betty Moore Foundation through grant GBMF5490 to the Ohio State University, and NSF grants AST-1515927 and AST-1908570. Development of ASAS-SN has been supported by NSF grant AST-0908816, the Mt. Cuba Astronomical Foundation, the Center for Cosmology and AstroParticle Physics at the Ohio State University, the Chinese Academy of Sciences South America Center for Astronomy (CAS-SACA), the Villum Foundation, and George Skestos. We acknowledge with thanks the observations from the AAVSO International Database contributed by observers worldwide and used in this research \citep[][]{aavsodb}.

The CSS survey is funded by the National Aeronautics and Space Administration under Grant No. NNG05GF22G was issued through the Science Mission Directorate Near-Earth Objects Observations Program. The Catalina Real-Time Transient Survey is supported by the U.S.~National Science Foundation under grants AST-0909182 and AST-1313422 \citep[][]{drake2009}. This paper has made use of up-to-date SMARTS optical/near-infrared light curves that are available at \url{www.astro.yale.edu/smarts/glast/home.php}. Data from the Steward Observatory spectropolarimetric monitoring project were used. This program is supported by Fermi Guest Investigator grants NNX08AW56G, NNX09AU10G, NNX12AO93G, and NNX15AU81G. This study makes use of data from the MOBPOL program conducted by S. Jorstad and A. Marscher at Boston University, and supported in part by the National Science Foundation under grant AST-1615796. Some of the data are based on observations collected at the Centro Astron\'{o}mico Hispano en Andaluc\'ia (CAHA); which is operated jointly by Junta de Andaluc\'{i}a and Consejo Superior de Investigaciones Cient\'{i}ficas (IAA-CSIC).

P.P. and M.A. acknowledge funding under NASA contract 80NSSC20K1562. J.O.S. acknowledges founding from the Istituto Nazionale di Fisica Nucleare Cap. U.1.01.01.01.009. S.B. acknowledges financial support by the European Research Council for the ERC Starting grant MessMapp under contract no. 949555. 

This work was supported by the European Research Council, ERC Starting grant \emph{MessMapp}, S.B. Principal Investigator, under contract no. 949555, and by the German Science Foundation DFG, research grant “Relativistic Jets in Active Galaxies” (FOR 5195, grant No. 443220636).

\bibliography{literature.bib} 
\bibliographystyle{aasjournal}

\begin{deluxetable*}{ccccccccc}[ht]
	\tablecaption{List of blazars studied, including their Fermi-LAT name, coordinates, AGN type, redshift, and association name. The sample consists in 6 flat-spectrum radio quasars (FSRQ) and 3 BL Lacertae (BL Lac). \label{tab:candidates_list}} 
	\tablewidth{0pt}
	\tablehead{
		\colhead{4FGL Source Name} &
		\colhead{RAJ2000} &
		\colhead{DecJ2000} &
		\colhead{Type} &
		\colhead{Redshift} & 
		\colhead{Association Name} & 
	}
	\startdata        
        4FGL J0222.6+4302 & 35.6696 & 43.0357 & BL Lac & 0.444 & 3C 66A \\
        4FGL J1104.4+3812 & 166.1187 & 38.2070 & BL Lac & 0.03 &  Mkn 421 \\
        4FGL J1224.9+2122 & 186.2277 & 21.3814 & FSRQ & 0.434 & 4C +21.35 \\
        4FGL J1229.0+0202 & 187.2675 & 2.0454 & FSRQ & 0.158 & 3C 273 \\
        4FGL J1256.1-0547 & 194.0415 & -5.7887 & FSRQ & 0.536 & 3C 279 \\
        4FGL J1653.8+3945 & 253.4738 & 39.7595 & BL Lac & 0.033 & Mkn 501 \\ 
        4FGL J1833.6$-$2103 & 278.4101 & -21.0574 & FSRQ & 2.507 & PKS 1830$-$211 \\ 
        4FGL J2232.6+1143 & 338.1525 & 11.7306 & FSRQ & 1.037 & CTA 102 \\
        4FGL J2253.9+1609 & 343.4963 & 16.1506 & FSRQ & 0.859 & 3C 454.3 \\
	\enddata
\end{deluxetable*}

\clearpage

\appendix \label{sec:appendix}
\renewcommand{\thesubsection}{\Alph{subsection}}
\renewcommand{\thefigure}{A\arabic{figure}}
\renewcommand{\thetable}{A\arabic{table}}

\subsection{Tables}
This section includes Table \ref{tab:results}, which presents the results of the periodicity analysis for the LC segments shown in the previous section.

\setcounter{table}{0}
\begin{deluxetable*}{ccccccccc}
	\tablecaption{Results of the characterization of the QPOs inferred. The table includes the temporal range of the QPO (MJD), and the parameters of the PSD model obtained from the PSD (``A'' and ``C'' ($rms^{2}/day^{-1}$)). Furthermore, the periods and uncertainties (top) are listed, along with their associated local significance (bottom). The final columns provide the global significance values, calculated by applying the trial correction described in $\S$\ref{sec:global_signifcance}.
	\label{tab:results}} 
	\tablewidth{0pt}
	\tablehead{
		\colhead{Source Name} &
            \colhead{LC Section} &
		\colhead{PSD Model} &
		\colhead{GLSP} &
		\colhead{SSA} &
        \colhead{Global Significance} &
        \colhead{Global Significance} &
		\\
            \colhead{} &
		\colhead{(MJD)} &
		\colhead{} &
		\colhead{} &
		\colhead{} &
        \colhead{(GLSP)} &
        \colhead{(SSA)} &
		\\
	}
	\startdata    
        \multirow{2}{*}{3C 66A} & 
        57790-58025 & 
        \makecell{$\alpha$=0.67$\pm$0.03 \\ $A$=0.063$\pm$0.001 \\ $C$=0.002$\pm$0.001} &
        $52.1^{\pm5.6}_{2.2\sigma}$ &
        $53.7^{\pm5.9}_{3.0\sigma}$ &
        $\approx$0$\sigma$ &
        $\approx$0$\sigma$ \\
        \hline 
	\multirow{2}{*}{3C 279} & 
        \makecell{\\ 57200-57650 \\ \\ \\ \\ 57790-58050  \\ \\ \\ \\ 58520-59060  \\ \\ } & 
        \makecell{
        \makecell{$\alpha$=0.47$\pm$0.03 \\ $A$=0.09$\pm$0.02 \\ $C$=0.016$\pm$0.008  \\ \\ } \\              \makecell{$\alpha$=0.99$\pm$0.07 \\ $A$=0.0011$\pm$0.0001 \\ $C$=0.055$\pm$0.002  \\ \\ } \\ 
        \makecell{$\alpha$=0.68$\pm$0.004 \\ $A$=0.018$\pm$0.004 \\ $C$=0.011$\pm$0.004  \\} } &
        \makecell{\\ $63.8^{\pm3.6}_{3.4\sigma}$ \\ \\ \\ \\ $54.2^{\pm5.3}_{2.8\sigma}$ \\ \\ \\ \\ $86.7^{\pm11.8}_{3.2\sigma}$ \\ \\ } &
        \makecell{\\ $63.7^{\pm5.9}_{3.6\sigma}$ \\ \\ \\ \\ $53.4^{\pm5.9}_{4.1\sigma}$ \\ \\ \\ \\ $85.7^{\pm4.2}_{4.0\sigma}$ \\ \\ } &
        \makecell{\\ 1$\sigma$ \\ \\ \\ \\ $\approx$0$\sigma$   \\ \\ \\ \\ $\approx$0$\sigma$  \\ \\ } &
        \makecell{\\ 1.4$\sigma$ \\ \\ \\ \\ 2.3$\sigma$   \\ \\ \\ \\ 2.3$\sigma$  \\ \\ } \\
        \hline
        \multirow{2}{*}{CTA 102} & 
        57170-57350 & 
        \makecell{$\alpha$=0.50$\pm$0.01 \\ $A$=0.011$\pm$0.002 \\ $C$=0.0015$\pm$0.0009} &
        $45.0^{\pm6.4}_{2.7\sigma}$ &
        $42.1^{\pm5.2}_{3.6\sigma}$ &
        $\approx$0$\sigma$ &
        1.4$\sigma$ \\
        \hline
        \multirow{2}{*}{3C 354.3} & 
        58040-58220 & 
        \makecell{$\alpha$=0.99$\pm$0.01 \\ $A$=0.0028$\pm$0.0002 \\ $C$=0.0011$\pm$0.0005} &
        $41.6^{\pm4.9}_{3.3\sigma}$ &
        $41.1^{\pm5.3}_{3.6\sigma}$ &
        1.0$\sigma$ &
        1.4$\sigma$ \\
        \hline
	\enddata
\end{deluxetable*}

\clearpage

\subsection{Light Curves}
This section reports the segments of the LCs included in Table \ref{tab:results} shown in Figure \ref{fig:parts_3c66a}, Figure \ref{fig:parts_3c279_1}, Figure \ref{fig:parts_3c279_2}, Figure \ref{fig:parts_3c279_3}, and Figure \ref{fig:parts_cta}.  

\setcounter{figure}{0}
\begin{figure*}
	\centering
	\includegraphics[scale=0.15]{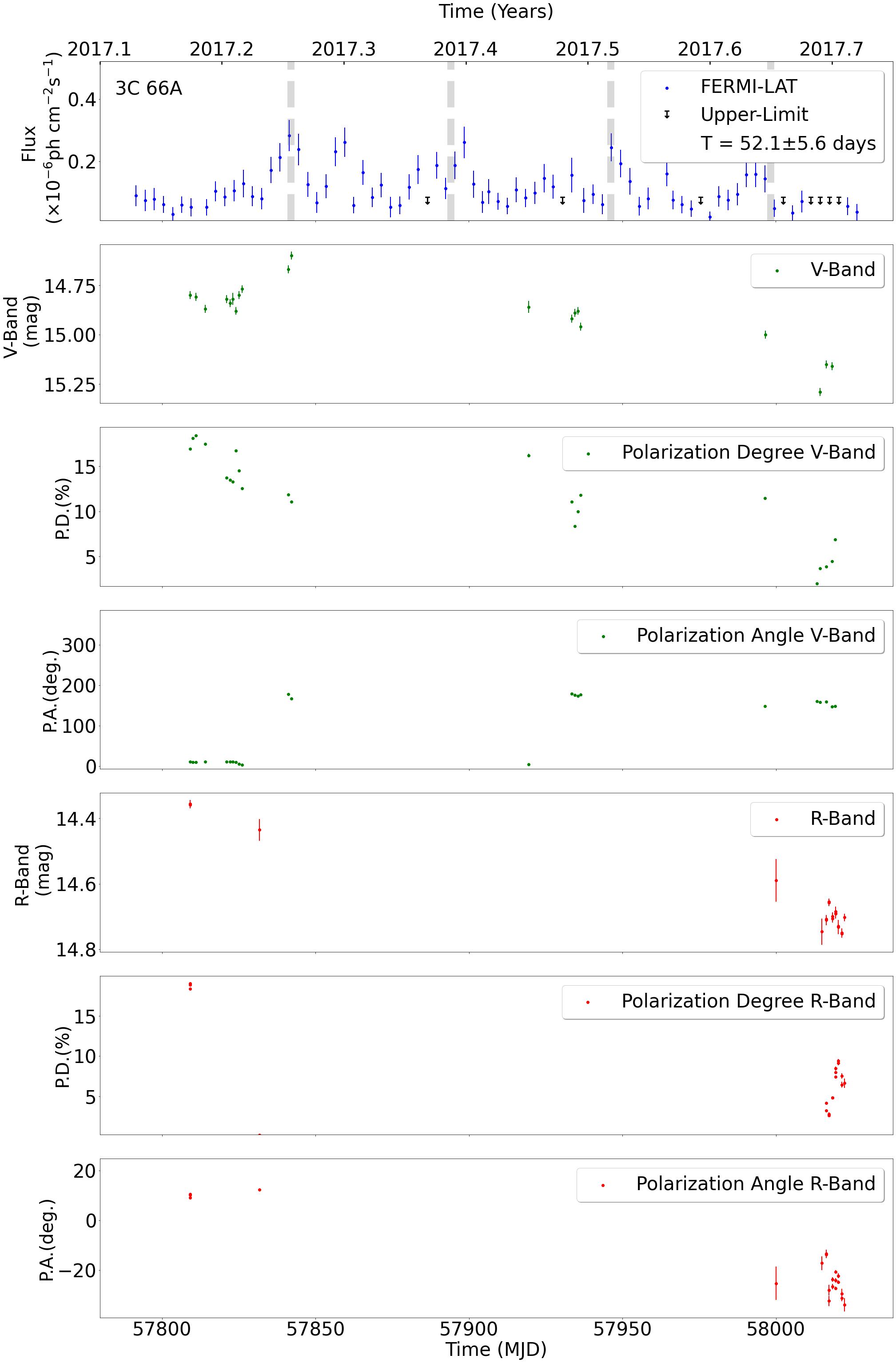}
	\caption{MWL light curves of 3C 66A for section 57800-58025. From top to bottom: \textit{Fermi}-LAT, V-Band, Polarization Degree (V-Band), Polarization Angle (V-Band), R-band, Polarization Degree (R-Band), Polarization Angle (R-Band) light curves.}
	\label{fig:parts_3c66a}
\end{figure*}

\begin{figure*}
	\centering
	\includegraphics[scale=0.15]{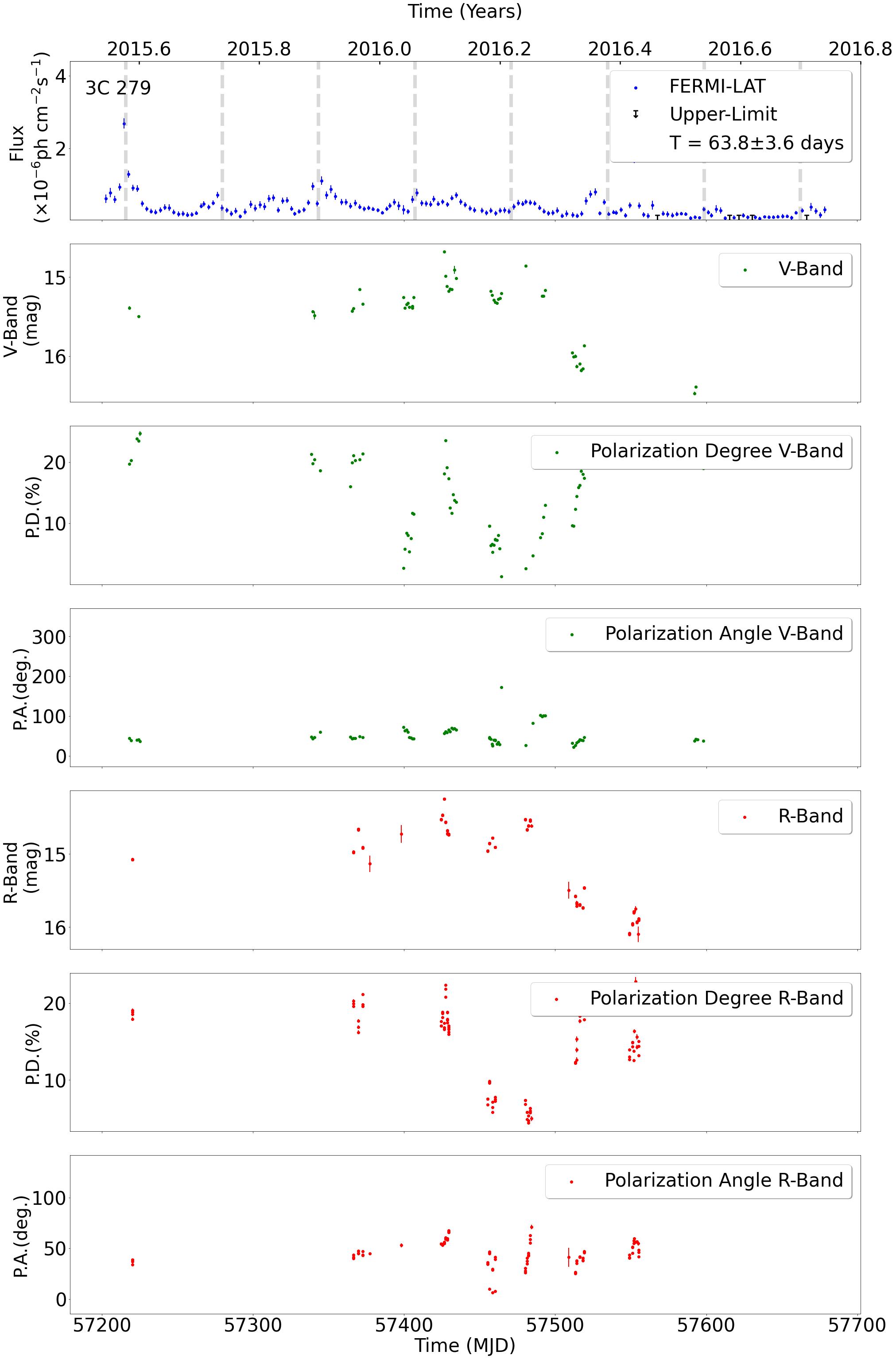}
	\caption{MWL light curves of 3C 279 for section 57200-57650. From top to bottom: \textit{Fermi}-LAT, V-Band, Polarization Degree (V-Band), Polarization Angle (V-Band), R-band, Polarization Degree (R-Band), and Polarization Angle (R-Band) light curves.}
	\label{fig:parts_3c279_1}
\end{figure*}

\begin{figure*}
	\centering
	\includegraphics[scale=0.15]{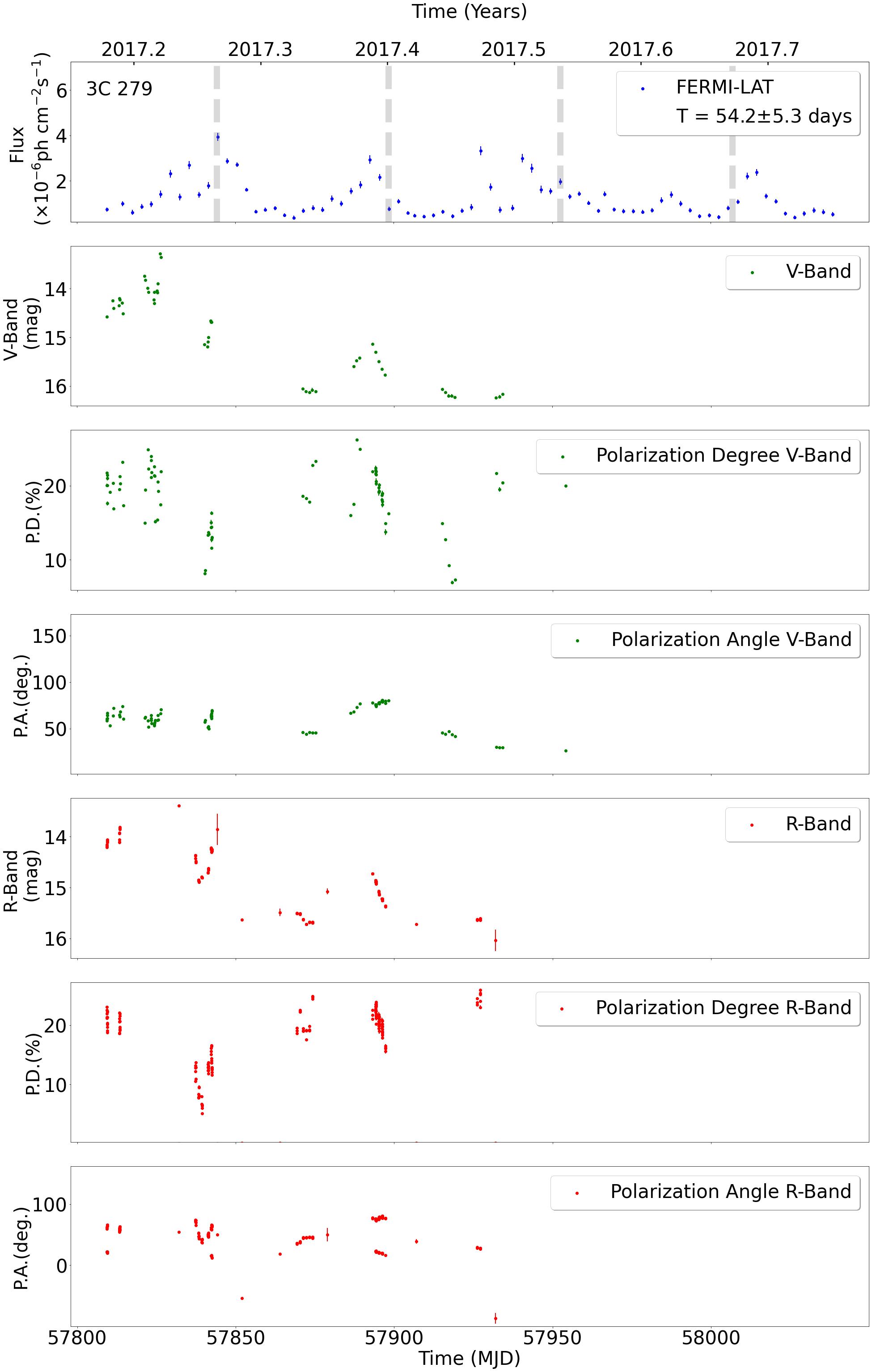}
	\caption{MWL light curves of 3C 279 for section 57800-58050. From top to bottom: \textit{Fermi}-LAT, V-Band, Polarization Degree (V-Band), Polarization Angle (V-Band), R-band, Polarization Degree (R-Band), and Polarization Angle (R-Band) light curves.}
	\label{fig:parts_3c279_2}
\end{figure*}

\begin{figure*}
	\centering
	\includegraphics[scale=0.15]{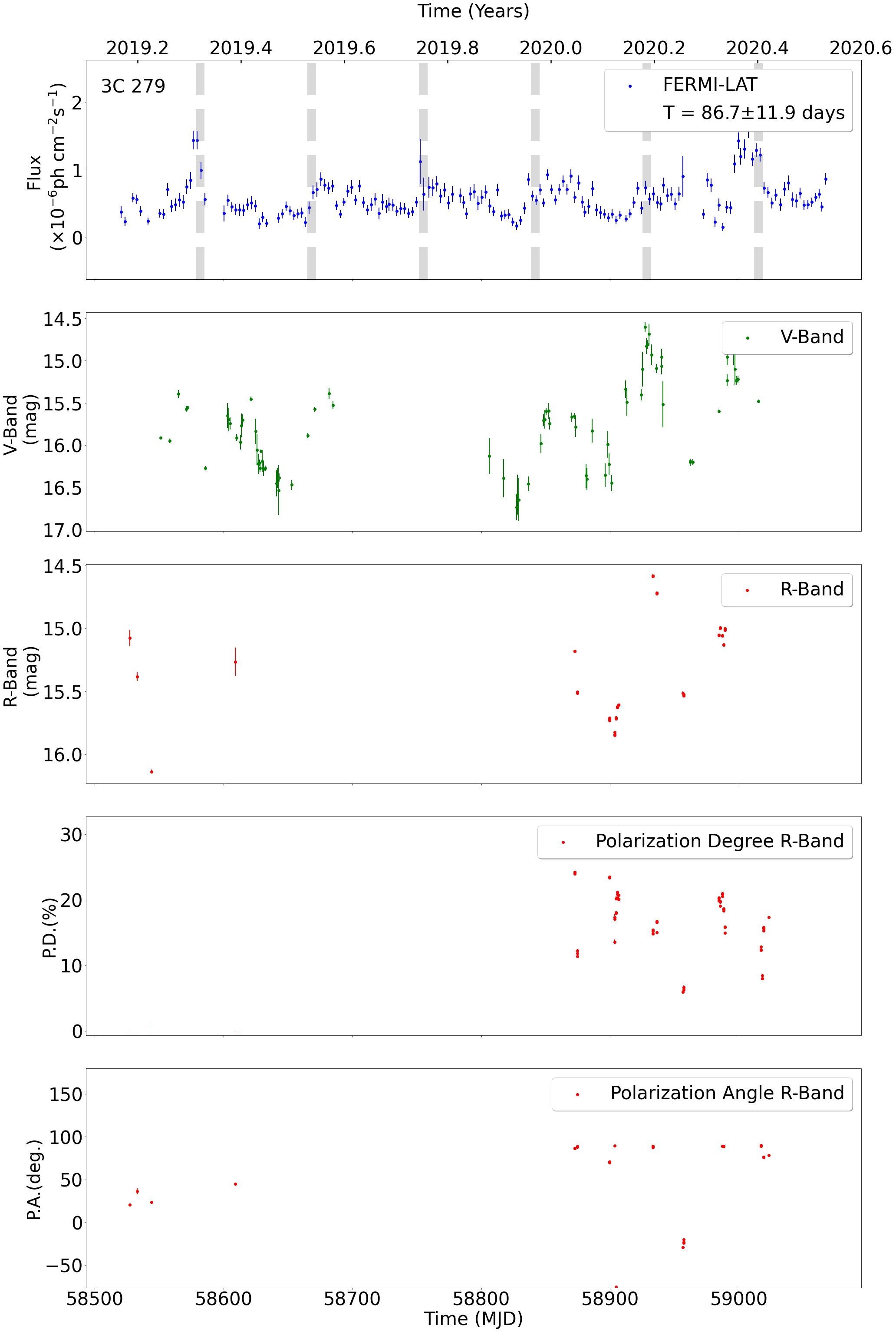}
	\caption{MWL light curves of 3C 279 for section 58520-59060. From top to bottom: \textit{Fermi}-LAT, V-Band, R-Band, Polarization Degree (R-Band), and Polarization Angle (R-Band) light curves.}
	\label{fig:parts_3c279_3}
\end{figure*}

\begin{figure*}
	\centering
	\includegraphics[scale=0.15]{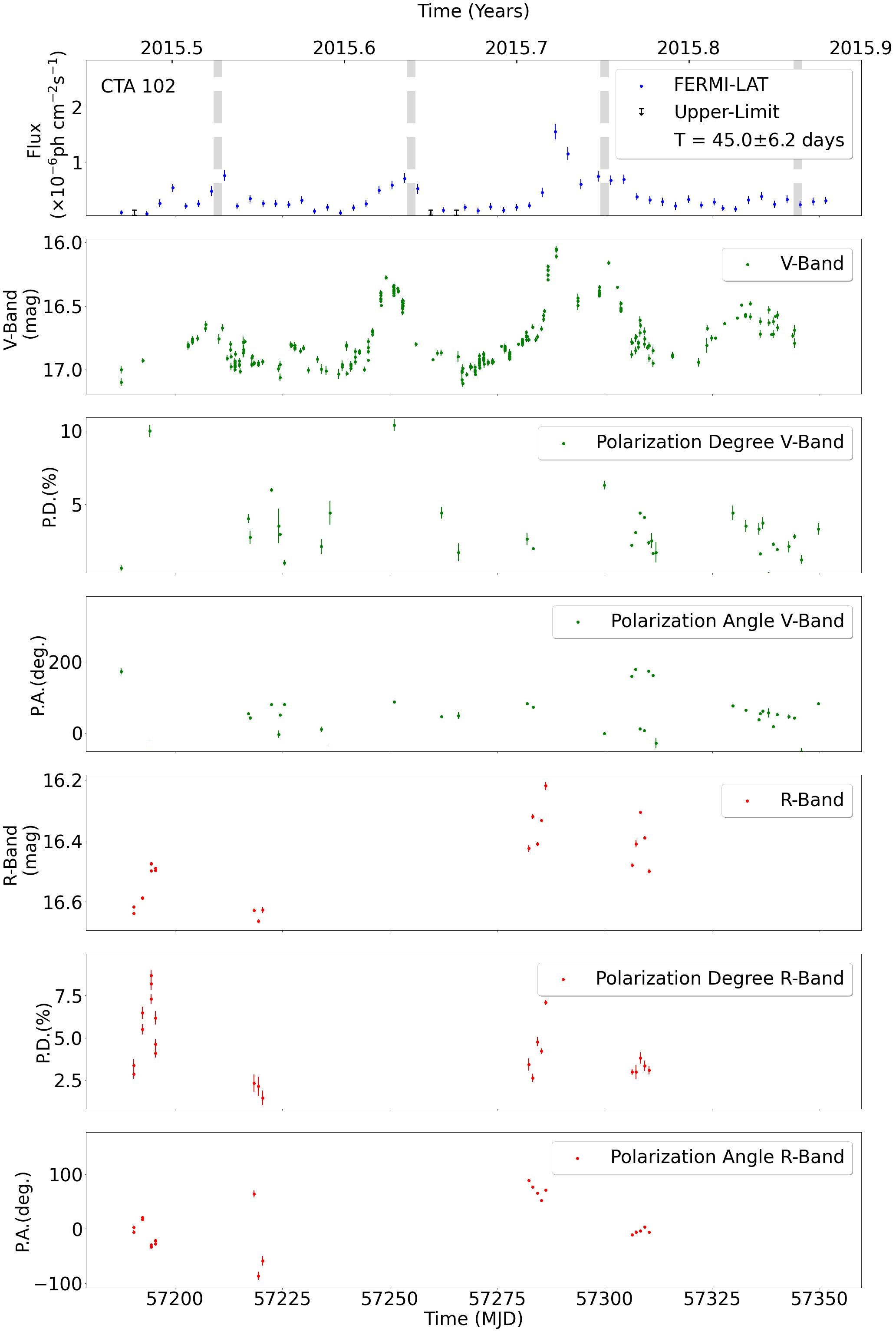}
	\caption{MWL light curves of CTA 102 for section 57170-57350. From top to bottom: \textit{Fermi}-LAT, V-Band, Polarization Degree (V-Band), Polarization Angle (V-Band), R-band, Polarization Degree (R-Band), and Polarization Angle (R-Band) light curves.}
	\label{fig:parts_cta}
\end{figure*}

\begin{figure*}
	\centering
	\includegraphics[scale=0.15]{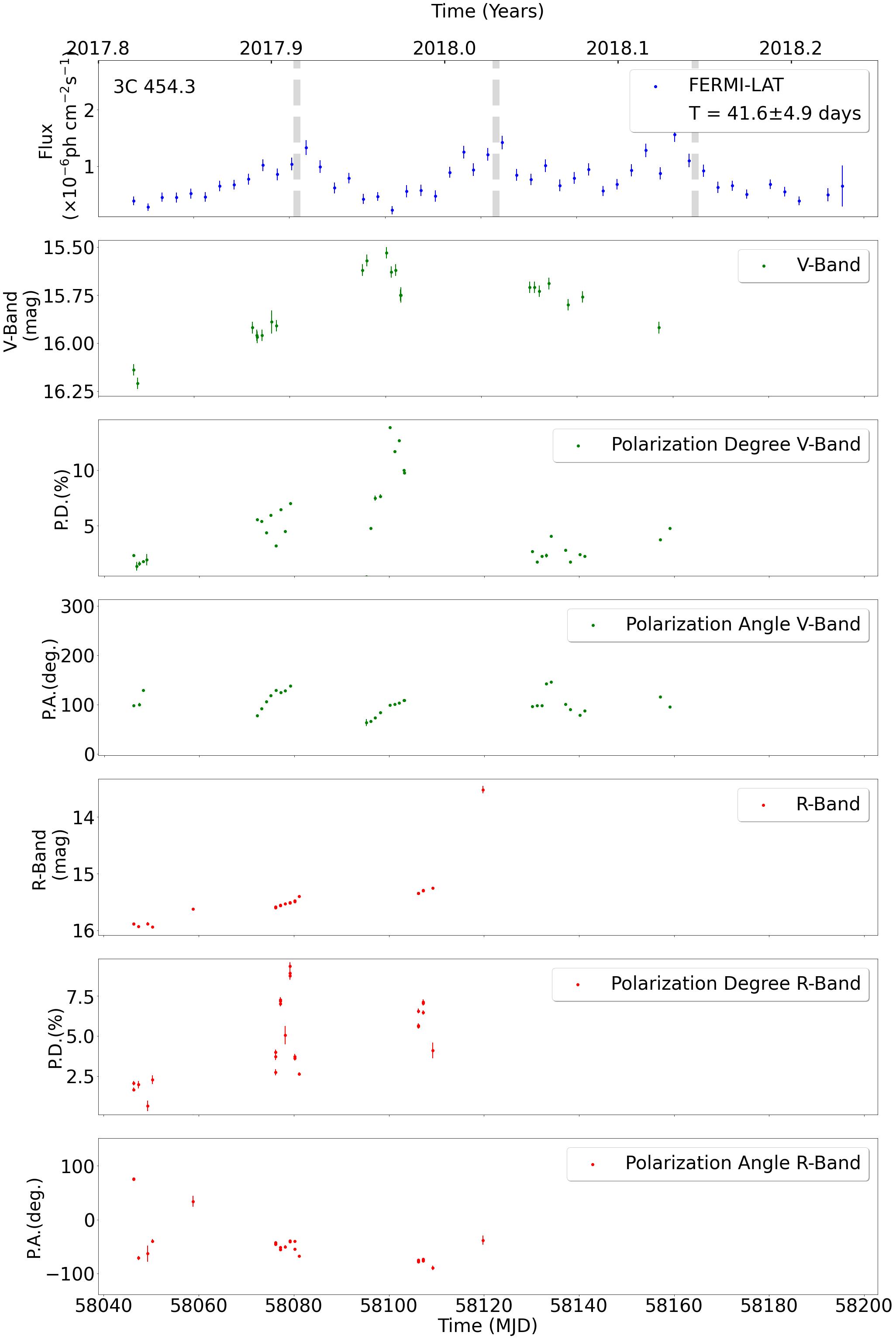}
	\caption{MWL light curves of 3C 354.3 for section 58040-58220. From top to bottom: \textit{Fermi}-LAT, V-Band, Polarization Degree (V-Band), Polarization Angle (V-Band), R-band, Polarization Degree (R-Band), and Polarization Angle (R-Band) light curves.}
	\label{fig:parts_3c454}
\end{figure*}

\clearpage

\end{document}